\documentclass[
  preprint,
  nofootinbib,
  aps,
  prb,
]{revtex4-2}

\usepackage{graphicx}
\usepackage{subcaption}
\usepackage{dcolumn}
\usepackage{bm}

\usepackage[top=15mm,bottom=20mm,left=15mm,right=15mm]{geometry}
\usepackage[doublespacing]{setspace}
\usepackage{amsmath}
\usepackage{amssymb}
\usepackage{bm,bbm}
\usepackage{subcaption}
\usepackage{physics,color}

\usepackage{caption}
\usepackage[
  setpagesize=false,
  bookmarks=true,
  bookmarksnumbered=true,
  colorlinks=true,
  citecolor=blue,
  linkcolor=blue,
  urlcolor=magenta,
  pdftitle={},
  pdfsubject={},
  pdfauthor={},
  pdfkeywords={}
]{hyperref}

\begin{document}

\preprint{APS/123-QED}

\title{DMFT analysis of Hopfield network with plasticity}

\author{Yoshinori Hara}
\email{hara-yoshinori793@g.ecc.u-tokyo.ac.jp}
\affiliation{
Department of Physics, The University of Tokyo, 7-3-1 Hongo, Tokyo 113-0033, Japan}.

\author{Yoshiyuki Kabashima}
\email{kaba@phys.s.u-tokyo.ac.jp}
\affiliation{
  Institute for Physics of Intelligence,
  The University of Tokyo, 7-3-1 Hongo, Tokyo 113-0033, Japan \\Trans-Scale Quantum Science Institute, The University of Tokyo, 7-3-1 Hongo, Tokyo 113-0033, Japan
}

\date{\today}

\begin{abstract}
  We study a fully connected Hopfield-type associative memory network with online activity-dependent synaptic plasticity, where neural states and synaptic couplings coevolve during retrieval. Using the generating-functional formalism, we derive a dynamical mean-field theory (DMFT) in the large-system limit with extensively many stored random patterns, and show that the many-body dynamics reduces to an effective single-site stochastic process with colored Gaussian crosstalk noise and delayed feedback terms. Numerical solutions of the DMFT equations agree well with direct simulations. We find that moderate plasticity enlarges the basin of attraction and increases the maximum retrievable memory load by generating a positive delayed feedback that stabilizes retrieval against crosstalk noise. However, excessively strong plasticity causes the network to imprint the imperfect initial cue itself, leading to spurious attractors and degraded retrieval performance. Consequently, an optimal plasticity strength emerges from the trade-off between memory stabilization and premature cue imprinting. These results extend the DMFT description of associative memory to networks with coevolving neural and synaptic dynamics.

\end{abstract}
\maketitle
\section{Introduction}

In classical theories of associative memory and working memory, synaptic couplings have generally been treated as fixed on the timescale of retrieval~\cite{Hopfield1982}, or at least as changing much more slowly than neuronal states~\cite{Amari1972}. This separation of learning and retrieval also underlies recent attempts to improve Hopfield-type networks by modifying the coupling matrix, such as unlearning and Daydreaming algorithms, where the couplings are optimized before retrieval and then treated as fixed interactions during the retrieval dynamics~\cite{Serricchio2025,Doi2026,takeuchi2026}. A typical example is the Hopfield-type associative memory model with nontrivial fixed points, in which couplings embedded by Hebbian learning act as quenched disorder and only the neurons evolve in time. Within this framework, statistical mechanics has achieved major success in describing macroscopic properties such as memory capacity, basins of attraction, and spin-glass-like behavior~\cite{Hopfield1982, PhysRevLett.55.1530}. Likewise, in the context of working memory, the view that information is maintained by persistent firing has long provided a standard theoretical foundation~\cite{MAJOR2004675}. In this sense, the separation of timescales between neurons and synapses has been an extremely powerful theoretical approximation.\par

In recent years, however, the validity of adopting this clear separation of timescales as a general principle has been reexamined. For example, studies of short-term synaptic plasticity and spike-timing-dependent plasticity have shown that synapses are not always static variables evolving sufficiently slowly compared with neural activity, and that at least some synaptic processes can change on timescales that overlap with neuronal computation~\cite{Tsodyks1997, Bi10464,Zucker2002}. Therefore, although the timescale separation assumed in classical theory remains a useful approximation, it can no longer be regarded as a self-evident premise for all phenomena of memory and learning.\par

This point is illustrated most clearly by a series of studies showing that synaptic efficacy can reflect recent neural activity in real time. The Tsodyks--Markram type of short-term plasticity formalized how synaptic responses change in a history-dependent manner on timescales ranging from a few milliseconds to several hundred milliseconds through depletion and recovery of available synaptic resources, thereby showing that synapses are not merely fixed transmission coefficients but possess their own internal dynamical states~\cite{Tsodyks1997}. Ref.~\cite{Gianluigi2008} theoretically proposed that working-memory contents can be retained through synaptic traces generated by short-term facilitation even in the absence of sustained high firing, providing a representative early example of what was later called ``activity-silent'' working memory. In addition, Ref.~\cite{Bittner2017} showed that hippocampal CA1 place fields can be formed in a single or a few trials through \emph{behavioral timescale synaptic plasticity} on the timescale of seconds, and Ref.~\cite{Milstein2021} showed that the same mechanism can bidirectionally reconfigure preexisting place fields. These results suggest that, at least in the hippocampus, synaptic updates do not depend solely on offline consolidation but can directly participate in the formation and updating of neural representations during behavior.\par

Against this experimental background, a theoretical framework that explicitly treats synaptic states as dynamical variables coupled to neural activity becomes compelling, at least in some situations, rather than one that strictly separates neuronal and synaptic states into distinct temporal hierarchies. Indeed, since Ref.~\cite{Tsodyks1998}, it has been discussed that recurrent networks with short-term depression and facilitation exhibit macroscopic dynamics qualitatively different from those of systems with fixed couplings. In associative-memory models as well, studies beginning with Ref.~\cite{Pantic2002} have examined how dynamic synapses modify memory phases, capacity, transitions between attractors, and stability. For example, Ref.~\cite{Mejias2009} analyzed capacity, Ref.~\cite{Otsubo2010} analyzed attractor switching, and Ref.~\cite{Katori2013} analyzed bifurcation structure. Furthermore, Ref.~\cite{Igarashi2010} showed that synaptic depression can generate new oscillatory states in more general recurrent and ring networks.
These studies show that activity-dependent changes in synaptic efficacy can alter memory phases, storage capacity, attractor stability, and switching dynamics. They include fully connected associative-memory networks as well as uniform and ring-attractor networks with dynamic synapses. However, they mainly analyze short-term depression and/or facilitation through low-dimensional or sublattice mean-field descriptions, often focusing on steady states, storage capacity, bifurcation structure, or attractor switching. They are therefore different from the present setting, where a fully connected Hopfield-type associative memory is supplemented by online Hebbian plastic couplings that coevolve with the retrieval dynamics.

What remains less explored is a DMFT description of fully connected
associative-memory models in which neuronal states and online plastic couplings
evolve with comparable time constants during retrieval.
For coupled neuronal--synaptic dynamics in random recurrent networks, Ref.~\cite{Clark2024} showed that synaptic dynamics not only modify collective dynamics but can, in some regimes, dominate them, and thereby provided a new phase diagram. In this sense, that work is the most direct precursor to the present study because it made clear on theoretical grounds that synapses should be treated not merely as slow background variables but as dynamical degrees of freedom on an equal footing with neurons. Meanwhile, as a recent reference system for dense-network DMFT specialized to associative memory, the analysis presented in Ref.~\cite{Kabashima_2026} for a nonplastic fully connected associative-memory model is important. Although their study does not address plasticity itself, it provides a methodological benchmark for the present work in terms of the construction of DMFT for fully connected associative memory, the numerical solution of the effective single-site process, and the comparison with direct simulations.

A particularly relevant recent development is the continuous-time study by Del Gaudio, Ghimenti, and Ganguli, who considered a Hopfield recurrent network with fixed Hebbian long-term couplings together with a rapidly evolving associative short-term plasticity matrix~\cite{DelGaudio2026}. Their model is closely related to the online Hebbian plasticity model studied here, apart from differences in notation, time discretization, and parameter scaling. By combining static cavity theory, DMFT, and finite-size simulations, they showed that a static fixed-point analysis captures only part of the retrieval behavior: while the cavity calculation predicts only a small shift of the equilibrium retrieval boundary, the
DMFT reveals a plastic-retrieval, or trampoline, mechanism that can stabilize transient retrieval beyond the static cavity boundary.

The present work addresses the same basic question from a complementary angle. Instead of focusing on the continuous-time formulation and the optimal plasticity timescale, we formulate a discrete-time generating-functional DMFT for retrieval dynamics. This gives a closed effective single-site process in which the crosstalk-induced Onsager reaction kernel and the plasticity-induced retarded self-interaction appear as distinct contributions. We then use this formulation to quantify how online plasticity reshapes the basin of attraction
to a retrieval state
as a function of initial overlap and memory load. In particular, we compute basin heat maps and the largest retrievable load after optimizing over plasticity strength, thereby making explicit the trade-off between stabilizing the target retrieval branch and prematurely imprinting the imperfect initial cue.

Taken together, what is currently needed is a theory that bridges, at the level of fully connected associative memory, the gap between classical fixed-synapse attractor memory and the online synaptic dynamics suggested by short-term plasticity and plasticity operating on behavioral timescales.
In this study, we consider an associative-memory network with fully connected Hebbian couplings supplemented by plastic couplings that are updated in real time according to neural activity, so that neuronal variables and synaptic variables coevolve with comparable time constants. In particular, we focus on situations in which synaptic updates proceed during the retrieval process itself, so that the coupling structure, conventionally regarded as fixed, is deformed on the fly. For this system, we construct a dynamical mean-field theory in the fully connected limit and compare numerical solutions based on the effective single-site process with direct simulations, thereby clarifying how plasticity affects memory retrieval and the basins of attractors. In this way, we aim to extend classical associative-memory theory, which assumes fixed synapses, to a more biologically plausible setting in which neurons and synapses evolve simultaneously.
This comparison also clarifies the scope of the present paper: rather than revisiting the continuous-time trampoline analysis of Ref.~\cite{DelGaudio2026}, we use the discrete-time generating-functional formulation introduced below to map how the retrieval basin depends jointly on initial overlap, memory load, and plasticity strength.

The remainder of this paper is organized as follows. In Sec.~II, we formulate the Hopfield network model with plasticity. In Sec.~III, we develop the dynamical mean-field theory for this model and derive the effective single-site process. In Sec.~IV, we compare the DMFT predictions obtained numerically from the effective process with direct simulations and assess how plasticity modifies retrieval dynamics. Finally, in Sec.~V, we summarize the results and discuss future directions.

\section{Model Setup}
We consider a variant of associative memory models consisting of $N$ neurons, whose state evolves in time according to the following equation:
\begin{align}
  \label{eq:dynamics}
  x^{t+1}_i &= x^t_i + \gamma \left( -x_i^t + \sum_{j \neq i} J_{ij}\,\phi(x_j^t) + \sum_{j \neq i} A_{ij}^t\,\phi(x_j^t) \right),\\
  s_i^t &= \mathrm{sign}(x_i^t).
\end{align}
Here, $\bm{x}^t = (x_1^t, \ldots, x_N^t)^\top \in \mathbb{R}^N$ denotes the state of the neurons at time $t$, $\phi(x) = \tanh(c x)$ is the activation function, and
$\bm{s} = (s_i)\in \{\pm 1\}^N$ stands for denotes pattern read out by the retrieval process.
The matrix $J_{ij}$ represents time-independent synaptic couplings, while $A^t = (A_{ij}^t) \in \mathbb{R}^{N \times N}$ denotes time-dependent couplings that evolve according to the Hebbian update rule
\begin{align}
  \label{eq:plasticity}
  A_{ij}^{t+1} = \rho A_{ij}^t + \frac{\beta}{N}\,\phi(x_i^t)\,\phi(x_j^t).
\end{align}
We impose the initial condition $A_{ij}^0=0$. By setting
\begin{align}
  \rho=1-\frac{\gamma}{\tau_A},\quad \beta=\frac{\gamma k_A}{\tau_A},
\end{align}
this model can be regarded as a discrete-time version of the continuous-time model with plastic synaptic couplings studied by Clark and Abbott~\cite{Clark2024}.
In their study, the fixed couplings were taken as $J_{ij}\sim \mathcal{N}(0,g^2/N)$. Here, by contrast, we consider couplings in which $P$ random binary patterns $\vb*{\xi}^\mu\in \qty{\pm 1}^N,\,(\mu=1,\ldots,P)$ are embedded via the Hebbian rule~\cite{Hopfield1982, PhysRevLett.55.1530}
\begin{align}
  J_{ij}=\frac{g}{N}\sum_{\mu=1}^P \xi_i^\mu\xi_j^\mu(1-\delta_{ij}).
\end{align}
That is, in the limit $c\to\infty$, $\beta\to 0$, and $\gamma\to 1$, the conventional Hopfield-network dynamics is recovered.\par
Our goal is to investigate how plasticity affects the memory-retrieval dynamics of the Hopfield network and to quantify, in terms of the plasticity-strength parameter $\beta$ or alternatively $k_A$, effects such as the expansion of the basin of attraction and the improvement of memory capacity.

\section{DMFT Analysis}

In this section, we first explain the logic of the derivation and then present only the main formulas needed in the remainder of the paper. The detailed algebra is deferred to Appendix~A--B. In what follows, we consider the limit $N\to\infty,\,P/N\to\alpha\sim \order{1}$~\cite{PhysRevLett.55.1530, PhysRevLett.61.259}.\par

The derivation proceeds in three steps. First, we encode the microscopic dynamics into a generating functional~\cite{Martin1973, Dominicis1978, Hertz_2017, Zou_2024}. Because the source fields couple to the full trajectory, this generating functional contains complete information about trajectory observables. Second, since the generating functional is normalized for each realization of the quenched disorder, the disorder-averaged trajectory statistics can be obtained by averaging the generating functional itself over the random patterns. Third, in the thermodynamic limit, the disorder-averaged generating functional is evaluated by the saddle-point method, which yields an effective single-site process without explicit spatial couplings.

\subsection{Generating Functional and Trajectory Observables}

First, it is useful to rewrite the plastic contribution as a retarded self-interaction. Solving Eq.~(\ref{eq:plasticity}) explicitly, following Ref.~\cite{Clark2024}, we obtain
\begin{align}
  A_{ij}^t=\frac{\beta}{N}\sum_{s=0}^{t-1}\rho^{t-1-s}\phi(x_i^s)\phi(x_j^s).
\end{align}

Accordingly, the local field generated by the plastic couplings can be written as
\begin{align}
  \sum_{j\neq i}A_{ij}^t\phi(x_j^t)
  =\sum_{s=0}^{t-1}\mathcal{K}_A(t,s)\phi(x_i^s)+\order{N^{-1}},
\end{align}
where
\begin{align}
  \mathcal{K}_A(t,s)=\beta\rho^{t-1-s}Q_N(t,s),
  \qquad
  Q_N(t,s)=\frac{1}{N}\sum_{i=1}^N\phi(x_i^t)\phi(x_i^s).
\end{align}

In the large system limit, $Q_N(t,s)$ is self-averaging and converges to the order parameter $Q(t,s)$~\cite{Dominicis1978, PhysRevLett.61.259, Hertz_2017}. Thus, the effect of plasticity appears as a delayed self-interaction acting on the past outputs of each neuron.

With this notation, Eq.~(\ref{eq:dynamics}) can be rewritten as
\begin{align}
  x_i^{t+1}=\mathcal{F}_i^t\qty[x_i^{0:s\le t}],
\end{align}
where
\begin{align}
  \mathcal{F}_i^t\qty[x_i^{0:s\le t}]
  =x_i^t+\gamma\qty(-x_i^t+\sum_{j\neq i}J_{ij}\phi(x_j^t)+\sum_{s=0}^{t-1}\mathcal{K}_A(t,s)\phi(x_i^s)).
\end{align}

Using this form, we define the generating functional for the trajectories as
\begin{align}
  \label{eq:generating_functional}
  Z(\theta)
  =\int \mathcal{D}x\,P_0(x^0)
  \prod_{t=0}^{T-1}\prod_{i=1}^N
  \delta\qty(x_i^{t+1}-\mathcal{F}_i^t\qty[x_i^{0:s\le t}])
  \exp\qty(i\sum_{t=1}^{T}\sum_{i=1}^N\theta_i^t x_i^t).
\end{align}

Here, $P_0(x^0)$ denotes the distribution of the initial condition. Since the source field $\theta$ couples to the variables at all times, $Z(\theta)$ generates arbitrary moments of the trajectories. In particular, under the normalization condition $Z(0)=1$, we have
\begin{align}
  \left.\frac{\partial Z(\theta)}{\partial(i\theta_i^t)}\right|_{\theta=0}
  &=\expval{x_i^t},\\
  \left.\frac{\partial^2 Z(\theta)}{\partial(i\theta_i^t)\partial(i\theta_j^s)}\right|_{\theta=0}
  &=\expval{x_i^t x_j^s},
\end{align}
and similarly for higher-order moments. Therefore, the generating functional contains complete information about the trajectory statistics.

\subsection{Why the Disorder Average of the Generating Functional Is Sufficient}

Our goal is to evaluate the trajectory statistics not for a fixed realization of patterns, but as an average over quenched random patterns to investigate the model's typical properties.
The key point is that the generating functional is normalized for each realization. Therefore, by differentiating the disorder-averaged generating functional, one directly obtains the disorder-averaged trajectory statistics.

Concretely, writing the pattern average as $\mathbb{E}_{\xi}[\cdots]$, we obtain
\begin{align}
  \mathbb{E}_{\xi}[Z(0)] &= 1,\\
  \left.\frac{\partial\mathbb{E}_{\xi}[Z(\theta)]}{\partial(i\theta_i^t)}\right|_{\theta=0}
  &=\mathbb{E}_{\xi}\qty[\expval{x_i^t}],\\
  \left.\frac{\partial^2\mathbb{E}_{\xi}[Z(\theta)]}{\partial(i\theta_i^t)\partial(i\theta_j^s)}\right|_{\theta=0}
  &=\mathbb{E}_{\xi}\qty[\expval{x_i^t x_j^s}],
\end{align}
with analogous relations for higher-order moments. Hence, instead of averaging each correlation function separately, it is sufficient to evaluate $\mathbb{E}_{\xi}[Z(\theta)]$. This is the central advantage of the generating-functional approach.

\subsection{Disorder-Averaged Generating Functional}

After carrying out the disorder average and saddle-point evaluation, as detailed in Appendix~B and Appendix~C, the averaged generating functional reduces to a product of identical single-site generating functionals without explicit spatial couplings.
\begin{align}
  \mathbb{E}_{\xi}[Z(\theta)]\simeq \prod_{i=1}^N Z_{\mathrm{eff}}[\theta_i],
\end{align}
where the generating functional for a representative site is given by
\begin{align}
  Z_{\mathrm{eff}}[\theta_i]
  &=\int \mathcal{D}x_i\,P_0(x_i^0)
  \int \mathcal{D}\eta_i\,\mathcal{N}(\eta_i\mid 0,C)
  \prod_{t=0}^{T-1}
  \delta\Biggl(
    x_i^{t+1}-x_i^t
    -\gamma\Big[
      -x_i^t+gm^t+\eta_i^t\\
      &+\sum_{s<t}\Lambda(t,s)\phi(x_i^s)
      +\sum_{s<t}\mathcal{K}_A(t,s)\phi(x_i^s)
    \Big]
  \Biggr)
  \exp\qty(i\sum_{t=1}^{T}\theta_i^t x_i^t).
\end{align}

This representation makes clear that, after the disorder average, the spatial couplings disappear and the problem reduces to a single-site process driven by colored Gaussian noise and memory kernels.

Accordingly, the representative trajectory obeys the effective single-site process
\begin{align}
  \label{eq:effective_single_site}
  x^{t+1}=x^t+\gamma\qty(-x^t+gm^t+\eta^t+\sum_{s<t}\Lambda(t,s)\phi(x^s)+\sum_{s<t}\mathcal{K}_A(t,s)\phi(x^s)),
\end{align}
where $\eta^t$ is a colored Gaussian noise with zero mean satisfying
\begin{align}
  \expval{\eta^t\eta^s}=C(t,s).
\end{align}

\subsection{Closed DMFT Equations}

The self-consistency conditions for the effective single-site process are
\begin{align}
  m^t&=\expval{\phi(x^t)},
  &Q(t,s)&=\expval{\phi(x^t)\phi(x^s)},\\
  \mathcal{K}_A(t,s)&=\beta\rho^{t-1-s}Q(t,s),
  &C(t,s)&=\alpha g^2(I-gG)^{-1}Q(I-gG)^{-\mathsf{T}}(t,s),\\
  &&\Lambda(t,s)&=\alpha g^2 G(I-gG)^{-1}(t,s).
\end{align}
Here, the response function $G(t,s)$ satisfies the causal condition $G(t,s)=0$ for $s\ge t$, and is updated numerically as
\begin{align}
  \label{eq:response_function_update}
  G(t,s)=
  \begin{cases}
    \displaystyle\sum_{w=0}^{T-1}\expval{\phi(x^t)\eta^w}\qty[C^{-1}_{<t}(w,s)], & s<t,\\
    0, & s\ge t.
  \end{cases}
\end{align}
Thus, the macroscopic dynamics of the fully connected associative-memory network with plasticity are described in closed form by an effective single-site process with colored Gaussian noise and two delayed feedback kernels. While the technical derivation is deferred to the appendices, this closed description is the main result of the present section.
\subsection{Schur complement}

Before turning to the numerical results, we briefly relate our numerical implementation to the continuous-time DMFT procedure of Ref.~\cite{DelGaudio2026}. In Ref.~\cite{DelGaudio2026}, the continuous-time equations are discretized on a finite time grid and solved as a global self-consistency problem for the two-time order parameters. The output correlation \(C_\phi(t,t')\), output response \(R_\phi(t,t')\), and condensed overlap \(m(t)\) are initialized on the full grid; the overlap response \(R_m\), noise covariance, and memory kernel are constructed from them; effective single-neuron trajectories are sampled; and the updated \(C_\phi\), \(R_\phi\), and \(m(t)\) are mixed back until convergence. The response \(R_\phi(t,t')\) is obtained by propagating tangent response equations along each effective trajectory. This scheme therefore solves the DMFT by repeated fixed-point iteration over the full time interval~\cite{PhysRevLett.61.259,DelGaudio2026}.

In contrast, the
formulation used here leads to a more explicitly causal implementation~\cite{Eissfeller_1992,Kabashima_2026}. At time \(t\), the construction only requires the past covariance block \(C_{<t}\), the sampled noise history, and the order parameters already obtained at earlier times. The response row \(G(t,s)\) for \(s<t\) is estimated from the Gaussian noise-response identity in Eq.~(25), after which \(Q\), \(G\), \(C\), \(\Lambda\), and \(\mathcal{K}_A\) are extended by one row and column. The next noise component \(\eta^t\) is then drawn from its conditional Gaussian distribution given the past noise history. Thus, rather than iterating a full two-time problem on a fixed grid, our implementation constructs the DMFT trajectory sequentially in time.

  This sequential construction also allows the inverse appearing in Eq.~(25) to be
  updated incrementally.  More generally, suppose that at time \(t\) the inverse
  of the past covariance block
  \[
    H_t \equiv C_{<t}^{-1}
  \]
  has already been computed.  When the next time point is appended, the enlarged
  block can be written as
  \[
    C_{<t+1}
    =
    \begin{pmatrix}
      C_{<t} & c_t \\
      c_t^{\top} & c_{tt}
    \end{pmatrix},
  \]
  where
  \[
    c_t = \bigl(C(0,t),\ldots,C(t-1,t)\bigr)^{\top},
    \qquad
    c_{tt}=C(t,t).
  \]
  If the Schur complement
  \[
    \Delta_t = c_{tt}-c_t^{\top}H_t c_t
  \]
  is nonzero, the inverse of the enlarged block is obtained from the block-inverse
  formula
  \[
    C_{<t+1}^{-1}
    =
    \begin{pmatrix}
      H_t + H_t c_t \Delta_t^{-1} c_t^{\top} H_t
      &
      -H_t c_t \Delta_t^{-1}
      \\
      -\Delta_t^{-1} c_t^{\top} H_t
      &
      \Delta_t^{-1}
    \end{pmatrix}.
  \]
  Thus, once \(H_t\) is available, the update to \(C_{<t+1}^{-1}\) only requires
  the matrix-vector product \(H_t c_t\) and a rank-one update.  The computational
  cost of this inverse update is therefore \(O(t^2)\), whereas recomputing the
  inverse of the enlarged \(t\times t\) block from scratch would cost \(O(t^3)\).

  In the response update, defining
  \[
    k_t(w)=\left\langle \phi(x^t)\eta^w \right\rangle,
    \qquad 0\leq w<t,
  \]
  Eq.~(25) can be evaluated as
  \[
    G(t,<t)=k_t^{\top}H_t .
  \]
  The same cached inverse is also used when drawing the next colored-noise
  component from its conditional Gaussian distribution~\cite{Rasmussen2005},
  \[
    \eta^t \mid \eta^{<t}
    \sim
    \mathcal{N}\left(
      c_t^{\top}H_t\eta^{<t},
      \,
      c_{tt}-c_t^{\top}H_t c_t
    \right).
  \]
  In this way, the causal row-by-row construction avoids repeated dense inversions
  of the growing covariance block.

When investigating transient phenomena such as relaxation to a retrieval state under given initial conditions, rather than a steady state, the causal approach of determining the next state solely from the past history shown here is natural.
  This approach is also computationally beneficial for the parameter scan below, because each parameter set can be advanced forward in time without an outer fixed-point iteration over the full two-time
correlation and response functions.

\section{Numerical Results}
In this section, we compare numerical solutions of the DMFT equations with the results of direct simulations~\cite{Eissfeller_1992, Kabashima_2026}. Throughout this section, we fix $\gamma=0.1,\,\tau_A=10,\, g=1.2,\, c=2$.

For a prescribed initial overlap $M^0$, the initial state was generated as follows. In the direct simulations, each spin was initialized independently according to
\begin{align}
  x_i^0=
  \begin{cases}
    \xi_i^1 & \text{with probability } (1+M^0)/2,\\
    -\xi_i^1 & \text{with probability } (1-M^0)/2.
  \end{cases}
\end{align}
In the DMFT calculations, a fraction $(1+M^0)/2$ of the trajectories was initialized from $x^0=1$, and the remaining fraction from $x^0=-1$.
\subsection{Comparison of the Time Evolution between DMFT and Direct Simulation}
We numerically solved Eqs.~(\ref{eq:effective_single_site}) and (\ref{eq:dynamics}) under different initial overlaps and pattern numbers $P=N\alpha$, and then compared the results.

  When solving the DMFT equations, however, the covariance block \(C_{<t}\)
  appearing in Eq.~(25) occasionally became nearly singular at sufficiently long
  times.  This occurs especially after the trajectory approaches a stationary
  retrieval state, where the colored-noise correlation becomes close to rank one.
  To stabilize the response regression, we therefore replaced the inverse in
  Eq.~(25) by a ridge-regularized inverse~\cite{Hoerl01021970, Ledoit2004},
  \[
    C_{<t}^{-1}
    \;\longrightarrow\;
    \left(C_{<t}+\lambda_g I_t\right)^{-1},
  \]
  where \(I_t\) is the \(t\times t\) identity matrix.  The same stabilization was
  also applied to the inverse used in the conditional sampling of the next
  colored-noise component, with an independent ridge parameter \(\lambda_{cg}\).
  Thus the two inverses used in the numerical implementation were
  \[
    H_t^{(g)}
    =
    \left(C_{<t}+\lambda_g I_t\right)^{-1},
    \qquad
    H_t^{(cg)}
    =
    \left(C_{<t}+\lambda_{cg} I_t\right)^{-1}.
  \]
  The sequential Schur-complement update described in Sec.~III.D was applied to
  these regularized covariance blocks.

\begin{figure}[htbp]
  \centering
  \includegraphics[width=\linewidth]{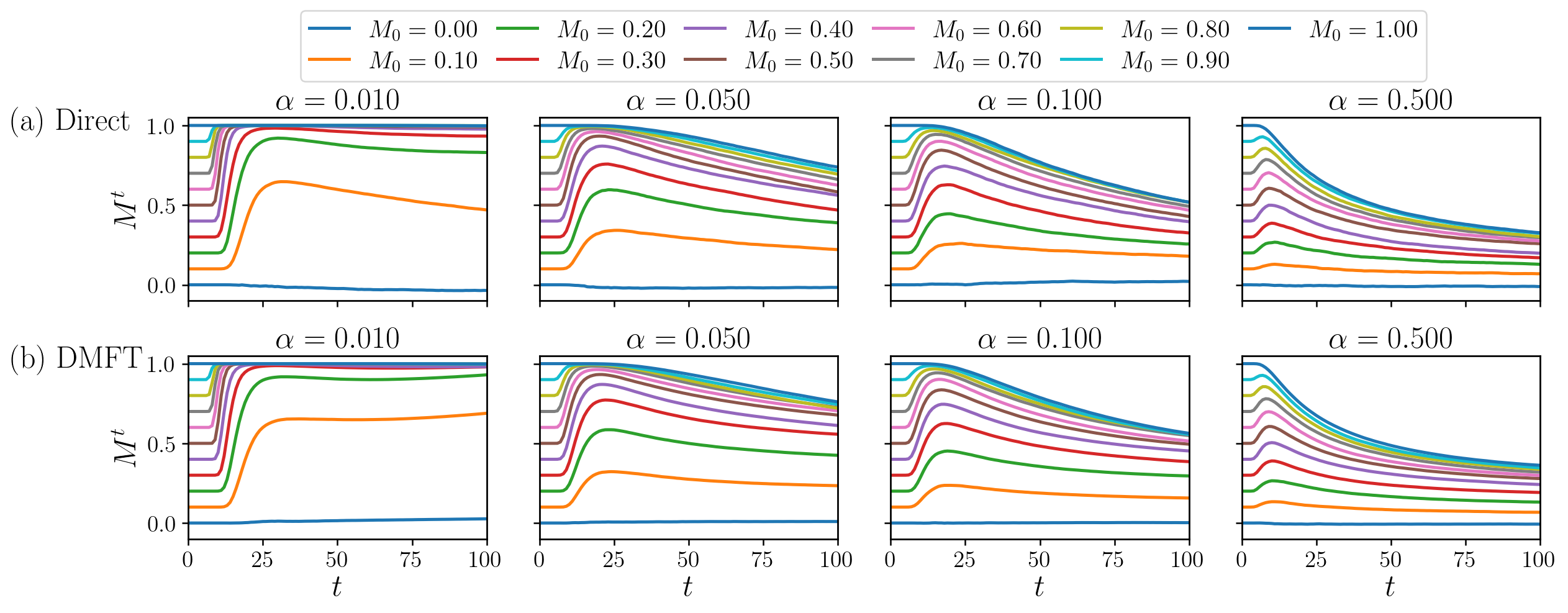}
  \caption{Time evolution of $M^t$ starting from different values of $M^0$ for $k_A=0$.}\label{fig:1}
\end{figure}

The results for the case $k_A=0$, namely in the absence of plasticity, are shown in Fig.~\ref{fig:1}. In the direct simulation we used $N=4000$, whereas the DMFT calculation used $N=10^6$ trajectories, with $\lambda_g=\lambda_{cg}=10^{-7}$.
\par
\begin{figure}[htbp]
  \centering
  \includegraphics[width=\linewidth]{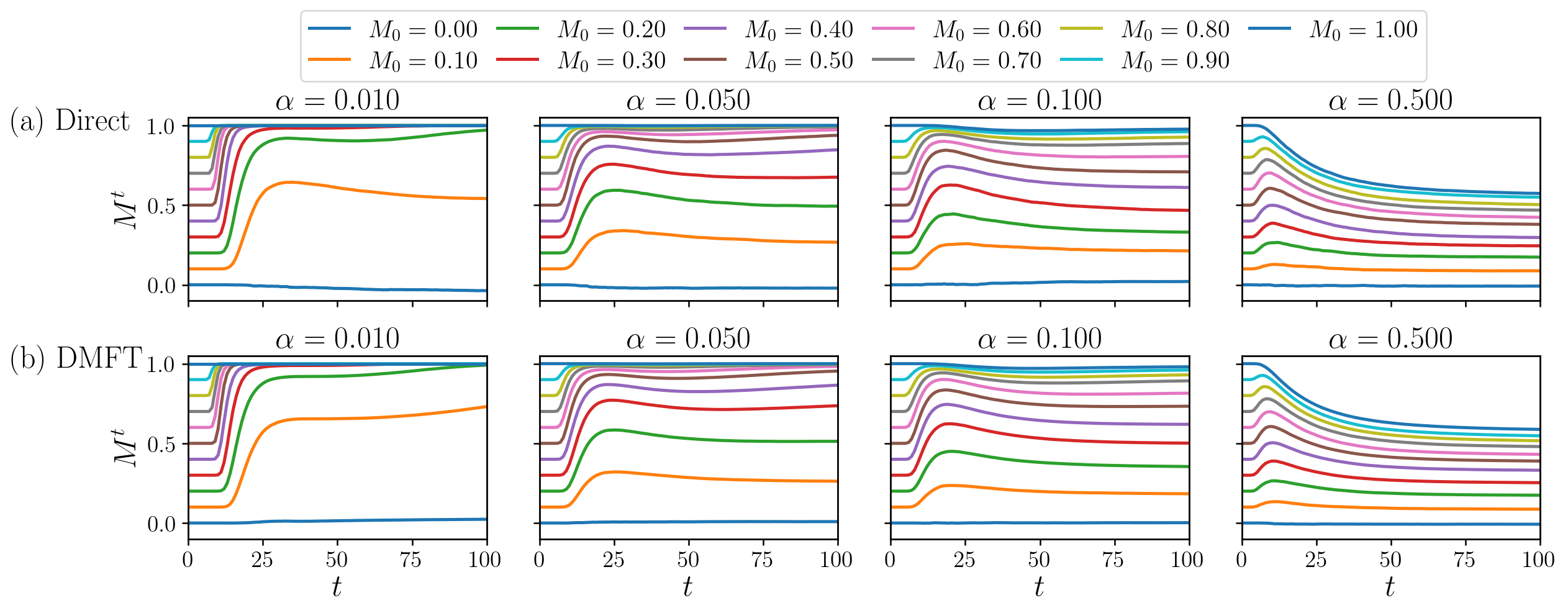}
  \caption{Time evolution of $M^t$ starting from different values of $M^0$ for $k_A=3.0$.}\label{fig:2}
\end{figure}

The results for $k_A=3.0$, in which the plasticity of the couplings is taken into account,  are shown in Fig.~\ref{fig:2}. All other parameters are the same.
\par
In both cases, the numerical solution of the DMFT equations agrees very well with the direct-simulation results. Moreover, compared with the case $k_A=0$, the system with $k_A=3.0$ reaches a final overlap close to $M^t=1$ even for larger values of $\alpha$. This indicates an improvement in memory capacity.
\subsection{Expansion of the Basin of Attraction}
\begin{figure}[htbp]
  \centering
  \includegraphics[width=\linewidth]{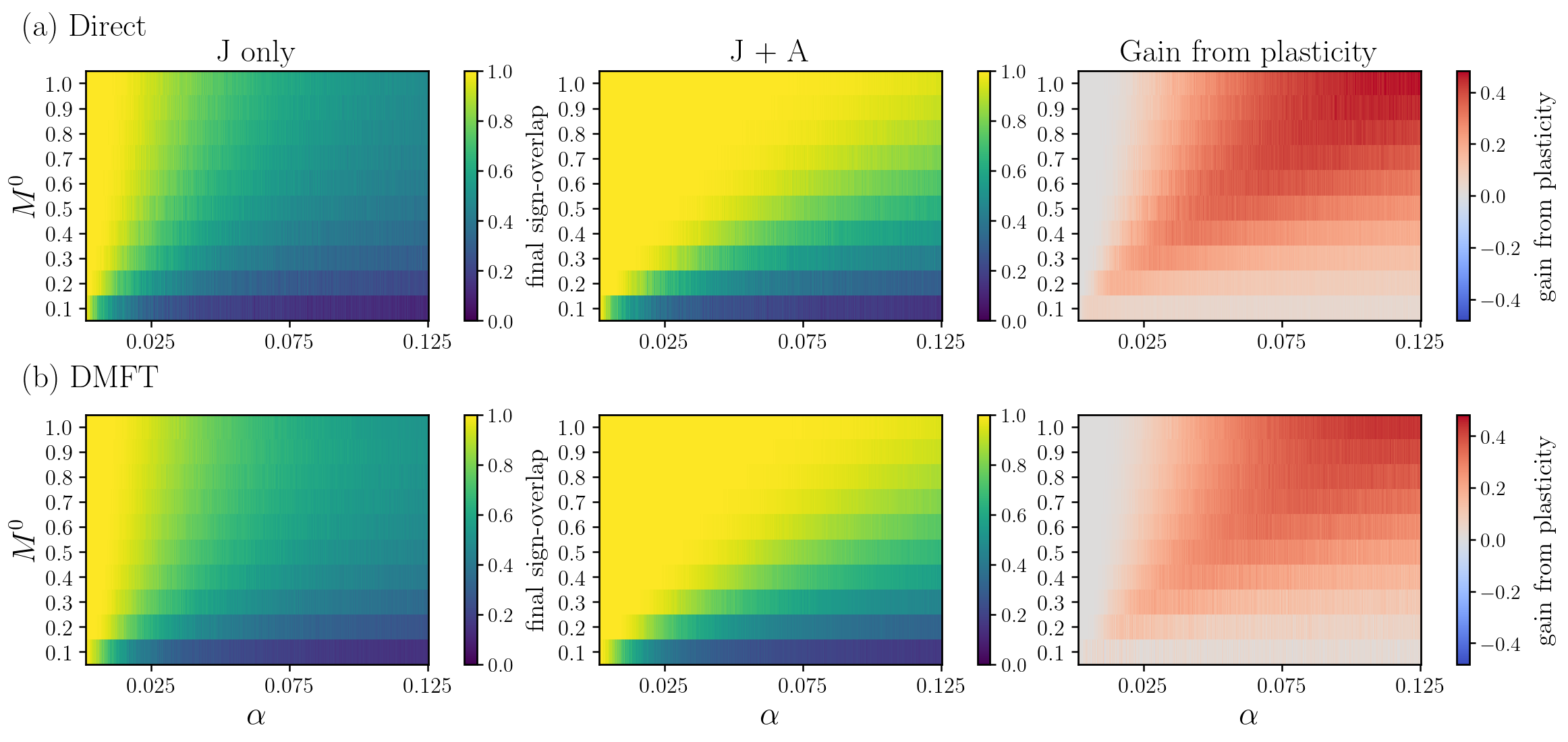}
  \caption{Comparison of heat maps of the final overlap obtained from direct simulation and DMFT for various initial overlaps $M^0$. In each row, the panels from left to right correspond to $k_A=0$, $k_A=3.0$, and their difference.}\label{fig:3}
\end{figure}

This basin-expansion question is closely connected to the dynamical stabilization identified in the continuous-time setting of Ref.~\cite{DelGaudio2026}. There, the emphasis was on plastic retrieval beyond the static cavity boundary and on the role of the plasticity timescale. Here, we ask a more quantitative question tailored to the present retrieval experiments: for a fixed initial overlap, how much does online plasticity enlarge the region in which the target memory is retrieved? To answer this, we scan both the memory load \(\alpha\) and the initial overlap \(M^0\), and compare the final overlap with and without plasticity.
Comparing Figs.~\ref{fig:1} and \ref{fig:2}, we see that even when starting from the same initial overlap and with the same memory load $\alpha$, the system with plasticity attains a higher final overlap. To quantify this effect, Fig.~\ref{fig:3} shows heat maps of the final overlap, with the initial overlap on the vertical axis and $\alpha$ on the horizontal axis. The final overlap was measured at time $T=100$, and the other parameters were the same as in the previous subsection.
\par
This figure shows that plasticity indeed contributes to an expansion of the basin of attraction~\cite{Pantic2002, Otsubo2010, Katori2013}, and that the DMFT qualitatively reproduces this expansion.\par
\begin{figure}[htbp]
  \centering
  \includegraphics[width=\linewidth]{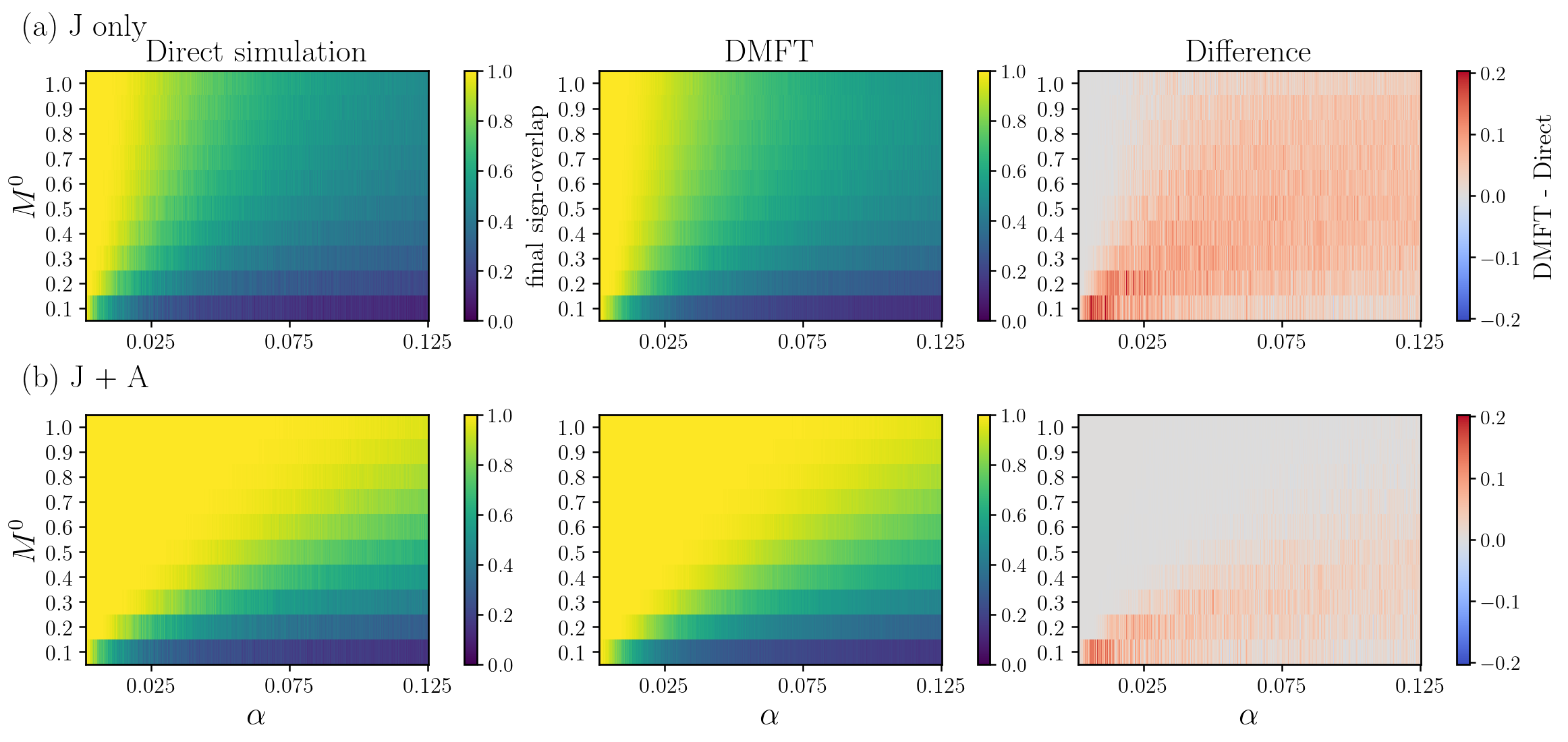}
  \caption{Residuals between direct simulation and DMFT for both $k_A=0$ and $k_A=3.0$.}\label{fig:4}
\end{figure}

For a more quantitative assessment, we next compare the residuals between the direct simulation and the DMFT results for both $k_A=0$ and $k_A=3.0$. The result is shown in Fig.~\ref{fig:4}.
This figure shows that, although the DMFT tends to slightly overestimate the results in the region of small $M^0$ for $k_A=0$, it nevertheless reproduces the direct-simulation results quantitatively overall.

\subsection{Converged Delayed Feedback and Fixed-Point Structure}

The results in the preceding subsections show that the DMFT effective process
reproduces both the time evolution and the basin expansion observed in direct
simulations. We now use the DMFT order parameters to examine the mechanism
behind this improvement. A useful reference point is the nonplastic
associative-memory DMFT studied
in Ref.~\cite{Kabashima_2026},
where the delayed feedback kernel \(\Lambda(t,s)\), corresponding to the
Onsager reaction term, plays a central role in determining the structure of the
converged retrieval state. In the present plastic model, however, the effective
single-site process contains two delayed self-feedback kernels: the Onsager
kernel \(\Lambda(t,s)\) generated by the crosstalk noise and the plastic kernel
\(K_A(t,s)\) generated by the online Hebbian plasticity. Therefore, the relevant
feedback acting on the converged state is not \(\Lambda(t,s)\) alone, but the
total delayed feedback
\begin{align}
  \Gamma(t,s)=\Lambda(t,s)+\mathcal{K}_A(t,s).
\end{align}
For the numerical experiment in this subsection, we
focus on a single retrieval condition rather than on a parameter sweep. We set
\(k_A=3.0\), and the memory load and initial overlap were fixed to
\begin{align}
  \alpha=0.1,\qquad M^0=0.8.
\end{align}
All other model parameters were the same as in the previous numerical
experiments.
The dynamics was run up to \(T=500\). The DMFT effective process was sampled
using \(N_{\rm path}=10^5\) trajectories. The response-function regression and
the conditional Gaussian sampling were both regularized by trace-scaled ridge
terms with
\begin{align}
  \lambda_g=\lambda_{cg}=10^{-7}.
\end{align}
The starred quantities below were evaluated by averaging over the last \(30\)
time steps. Direct simulations were not used in this diagnostic calculation;
the purpose here is to decompose the already validated DMFT effective process
into its Onsager and plastic feedback components.

To quantify the strength of these feedback terms, we define their integrated
forms as
\begin{align}
  \Lambda(t)
  &=
  \sum_{s<t}\Lambda(t,s),
  \\
  K_A(t)
  &=
  \sum_{s<t}\mathcal{K}_A(t,s),
  \\
  \Gamma(t)
  &=
  \Lambda(t)+K_A(t).
\end{align}
This integrated quantity is especially useful because individual components of
\(\Lambda(t,s)\) can fluctuate strongly after convergence due to the
near-singularity of the noise correlation matrix, whereas their sum gives the
net feedback acting on the effective single-site dynamics.

\begin{figure}[t]
  \centering
  \includegraphics[width=\linewidth]{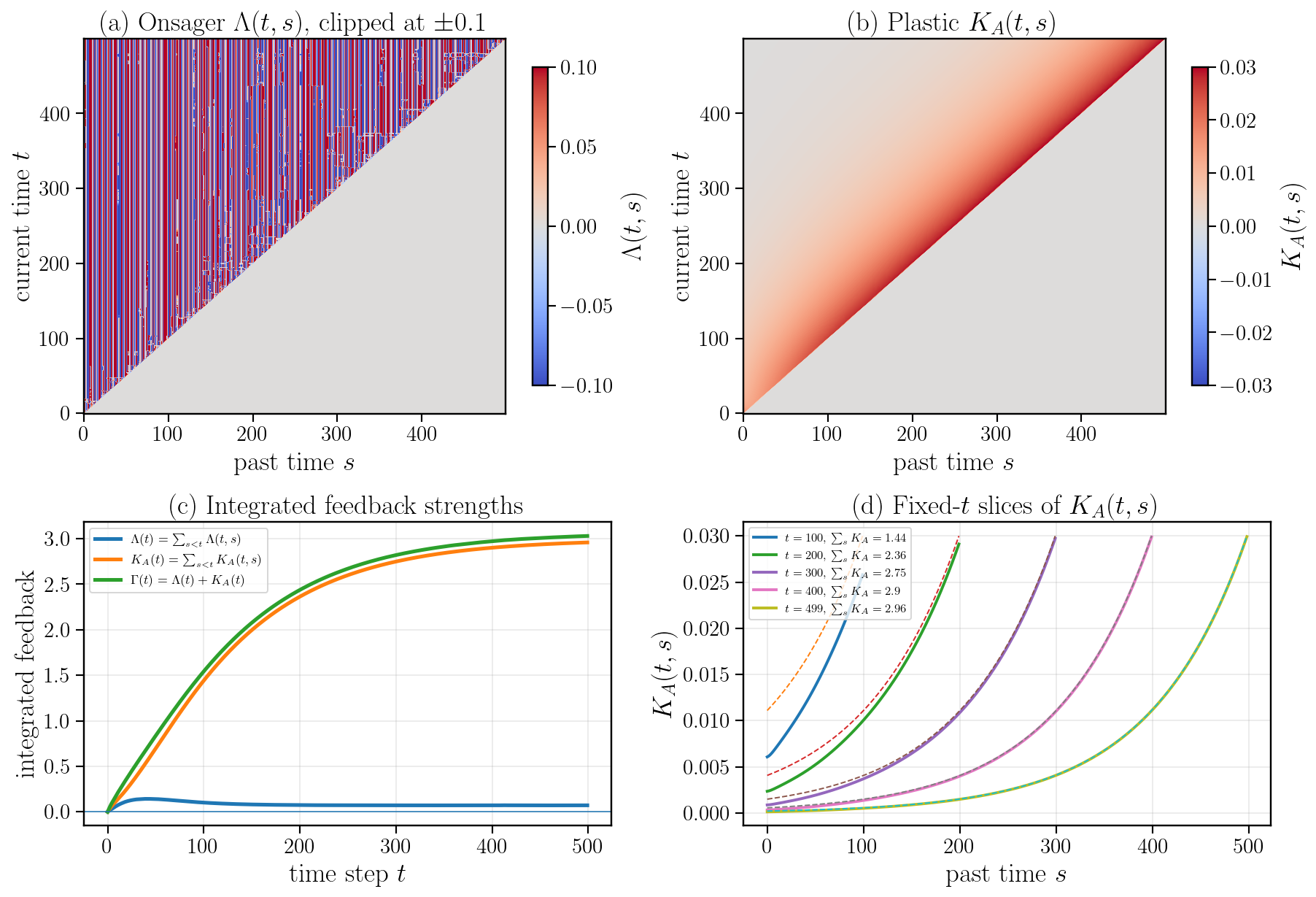}
  \caption{
    Structure and integrated strength of the delayed feedback kernels in the
    converged retrieval state.
    (a) Clipped heat map of the Onsager kernel \(\Lambda(t,s)\).
    The strong sign-changing fluctuations are mainly numerical fluctuations
    amplified by the inversion of the nearly singular noise correlation matrix.
    (b) Clipped heat map of the plastic kernel \(K_A(t,s)\), which shows a
    smooth exponentially weighted memory structure.
    (c) Time evolution of the integrated Onsager feedback \(\Lambda(t)\), the
    integrated plastic feedback \(K_A(t)\), and their sum \(\Gamma(t)\).
    The total feedback is dominated by the plastic contribution in the
    converged retrieval state.
    (d) Fixed-\(t\) slices of \(K_A(t,s)\). The dashed curves show
    \(\beta\rho^{t-1-s}\), which is obtained by assuming \(Q(t,s)\simeq 1\).
  }
  \label{fig:feedback_kernel_summary}
\end{figure}

Figure~\ref{fig:feedback_kernel_summary}(c) shows the time evolution of these
integrated feedbacks. The Onsager contribution remains small in the converged
state, while the plastic contribution grows slowly and saturates at an
\(O(k_A)\) value. In the present run, the tail-averaged values were
approximately
\begin{align}
  &\Lambda_*
  \simeq
  7.1\times 10^{-2},
  \\
  &K_{A,*}
  \simeq
  2.95,
  \\
  &\Gamma_*
  =
  \Lambda_*+K_{A,*}
  \simeq
  3.02.
\end{align}
Thus, the total feedback is dominated almost entirely by the plastic component
\(K_{A,*}\). This observation indicates that, in the plastic model, the
converged retrieval state is controlled primarily by the total delayed feedback
produced by the coupling plasticity
rather than by the Onsager reaction term alone.

This behavior is also consistent with the form of the plastic kernel. When the
system is close to a retrieval state, \(Q(t,s)\simeq 1\) and
\(\phi(x^t)\simeq\phi(x^s)\). In this regime, the plastic contribution
approximately becomes
\begin{align}
  \sum_{s<t}\beta\rho^{t-1-s}Q(t,s)\phi(x^s)
  &\simeq
  \sum_{s<t}\beta\rho^{t-1-s}\phi(x^t)
  \\
  &=
  \frac{\beta}{1-\rho}\qty(1-\rho^t)\phi(x^t)
  \\
  &\simeq
  k_A\phi(x^t),
\end{align}
and therefore acts as a positive self-reinforcing feedback. The saturation of
\(K_A(t)\) around \(k_A\) in Fig.~\ref{fig:feedback_kernel_summary}(c) is
consistent with this approximation.

We next examine the structure of the feedback kernels themselves.
Figures~\ref{fig:feedback_kernel_summary}(a) and
\ref{fig:feedback_kernel_summary}(b) show clipped heat maps of
\(\Lambda(t,s)\) and \(\mathcal{K}_A(t,s)\), respectively. The two kernels have clearly
different structures. The plastic kernel shows a smooth triangular profile,
reflecting the exponentially weighted memory trace
\(\beta\rho^{t-1-s}\). By contrast, the Onsager kernel exhibits strong
sign-changing fluctuations even after clipping. This does not mean that the
net Onsager feedback is large. Rather, as in the analysis of Ref.~\cite{Kabashima_2026}, once the dynamics approaches a retrieval state, the noise correlation
matrix becomes nearly rank-one. The inversion of this nearly singular matrix
can amplify Monte Carlo fluctuations in the response estimate, which then
appear as large fluctuations in the individual components of \(\Lambda(t,s)\).
Therefore, the physically relevant quantity is not each raw element of
\(\Lambda(t,s)\), but the row-integrated feedback \(\Lambda(t)\), where
positive and negative fluctuations largely cancel.

This point is further illustrated by the comparison between
Figs.~\ref{fig:feedback_kernel_summary}(c) and
\ref{fig:feedback_kernel_summary}(d). Although the raw Onsager kernel is
difficult to interpret element by element, its integrated value remains small
and stable at late times. In contrast, the plastic kernel follows the expected
exponential profile
\begin{align}
  K_A(t,s)
  \simeq
  \beta\rho^{t-1-s},
\end{align}
as shown by the fixed-\(t\) slices in
Fig.~\ref{fig:feedback_kernel_summary}(d). Its row sum approaches \(O(k_A)\).
Thus, the integrated quantities show a clear separation:
\begin{align}
  |\Lambda_*| \ll K_{A,*},
  \qquad
  \Gamma_* \simeq K_{A,*}.
\end{align}

We next examine whether the converged feedback explains the observed
single-site fixed point. Let
\begin{align}
  u_*
  &=
  \phi(x_*),
  \\
  h_*
  &=
  g m_*+\eta_*,
\end{align}
where \(h_*\) is the effective signal plus crosstalk noise. The factor \(g\) is
included because the signal term in the present effective process is \(g m^t\).

If the integrated feedback has converged to \(\Gamma_*\), the fixed-point
relation becomes
\begin{align}
  u_*
  =
  \phi(h_*+\Gamma_*u_*).
\end{align}
For comparison, we also consider the Onsager-only relation
\begin{align}
  u_*
  =
  \phi(h_*+\Lambda_*u_*).
\end{align}

\begin{figure}[t]
  \centering

  \begin{subfigure}{0.48\linewidth}
    \centering
    \includegraphics[width=\linewidth]{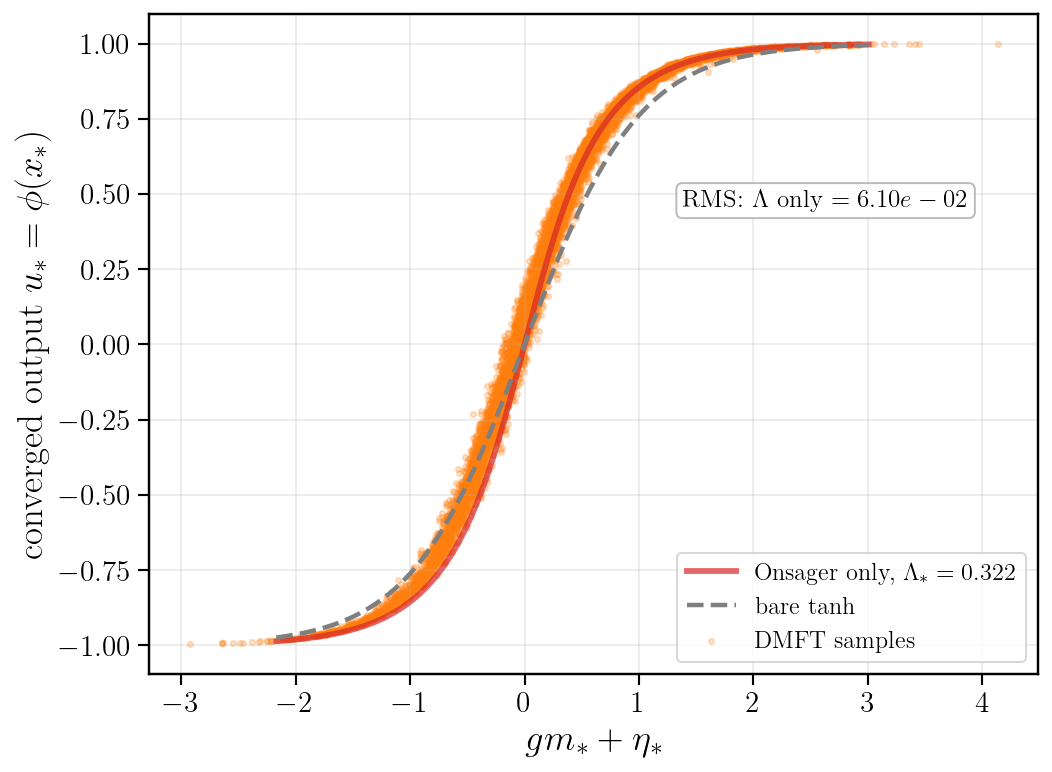}
    \caption*{(a)}
  \end{subfigure}
  \hfill
  \begin{subfigure}{0.48\linewidth}
    \centering
    \includegraphics[width=\linewidth]{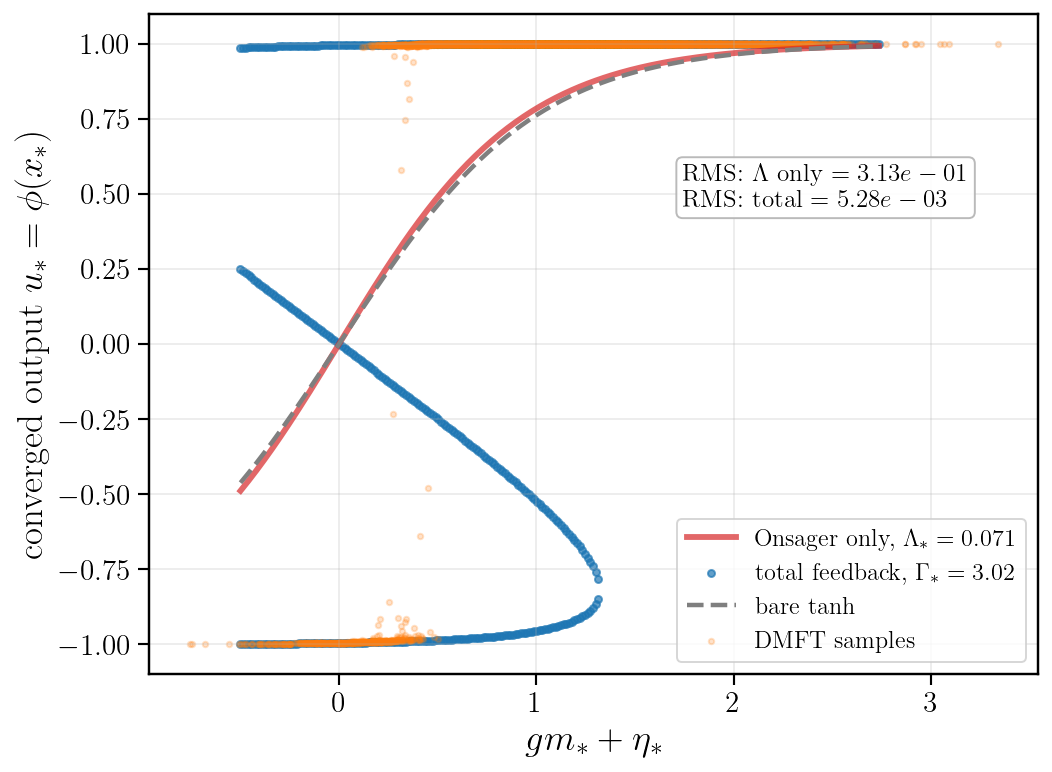}
    \caption*{(b)}
  \end{subfigure}

  \caption{
    Converged input-output relations of the DMFT effective process.
    The horizontal axis is \(h_*=g m_*+\eta_*\), and the vertical axis is
    \(u_*=\phi(x_*)\).
    (a) No-plasticity control. Since \(K_A=0\), the total feedback is identical
    to the Onsager feedback, and the effective relation remains close to a
    single-valued tanh curve.
    (b) Plastic case with \(k_A=3.0\). The Onsager-only relation fails to
    reproduce the converged samples, whereas the total-feedback relation
    \(u_*=\phi(h_*+\Gamma_*u_*)\) accurately captures them.
  }
  \label{fig:fixed_point_comparison}
\end{figure}

Figure~\ref{fig:fixed_point_comparison}(a) shows the no-plasticity control.
Since \(K_A=0\), the total feedback is identical to the Onsager feedback,
\begin{align}
  \Gamma_*=\Lambda_*.
\end{align}
In this case, \(\Lambda_*\simeq 0.32\), and the effective fixed-point relation
remains close to a single-valued tanh curve. Thus, without plasticity, the
final output of each site is mainly determined by the instantaneous effective
input \(h_*=g m_*+\eta_*\), with only a moderate Onsager correction.

Figure~\ref{fig:fixed_point_comparison}(b) shows the plastic case. The
Onsager-only relation is close to the bare transfer function because
\(\Lambda_*\simeq 7.1\times 10^{-2}\) is small, and it does not reproduce the
converged DMFT samples. By contrast, the curve obtained using the total
feedback \(\Gamma_*=\Lambda_*+K_{A,*}\) accurately captures the observed
input-output relation. In the present run, the root-mean-square residual
decreases from approximately
\begin{align}
  3.13\times 10^{-1}
\end{align}
for the Onsager-only relation to
\begin{align}
  5.28\times 10^{-3}
\end{align}
for the total-feedback relation. Thus, the fixed-point structure of the plastic
retrieval state is determined by the total delayed feedback.

An important consequence is that even with the monotonic transfer function
\(\phi(x)=\tanh(x)\), the effective fixed-point equation can have multiple
algebraic roots when the positive feedback is sufficiently strong. In the
present case, \(\Gamma_*\simeq 3.02>1\), so the scalar equation
\begin{align}
  u
  =
  \tanh\qty(h+\Gamma_*u)
\end{align}
can become multivalued. This multivalued structure is not caused by a
nonmonotonic transfer function, but by the large positive self-feedback
generated dynamically by plasticity.

However, the existence of multiple roots itself is not the mechanism that
improves retrieval. The important point is that, once the retrieval trajectory
has selected the branch corresponding to the target memory, the positive
plastic feedback stabilizes that branch against crosstalk noise. In other
words, plasticity changes the effective input-output relation from a smooth
instantaneous tanh response to a history-dependent branch-selection mechanism.
This makes the correctly selected retrieval branch more robust.

This interpretation also resolves the apparent difference between intermediate
and long observation times. At an intermediate time such as \(T=200\), the
plastic feedback has already become strong enough to generate multiple roots,
but the plastic trace has not fully saturated:
\begin{align}
  \Gamma_*(T=200)
  \simeq
  2.3.
\end{align}
At the longer time \(T=500\), the feedback has grown further:
\begin{align}
  \Gamma_*(T=500)
  \simeq
  3.0.
\end{align}
The final DMFT samples therefore relax onto the stable saturated branches. The
absence of samples on the intermediate branch does not mean that the
multibranch structure has disappeared; it indicates that the intermediate
branch is dynamically unstable and that the history of the DMFT dynamics
selects one of the stable branches.

Taken together, these results show that plasticity improves retrieval by a
mechanism distinct from a simple increase of the static signal gain. In the
nonplastic system, the converged response is governed by the Onsager feedback
\(\Lambda_*\), which remains moderate and yields a nearly single-valued tanh
relation. In the plastic system, online Hebbian updates generate an additional
trajectory-dependent delayed self-feedback \(\mathcal{K}_A(t,s)\). This feedback
dominates the converged total feedback \(\Gamma_*\), locks in the correctly
selected retrieval branch, and thereby makes the retrieval state more robust
against crosstalk noise. The same mechanism also suggests why excessively large
plasticity can become harmful: if the wrong branch associated with the initial
cue is selected too early, plastic feedback stabilizes that spurious state
instead of the target memory. This trade-off motivates the search for the
optimal plasticity strength in the next subsection.

\subsection{Estimation of the Optimal Plasticity Strength}
\label{subsec:optimal_plasticity}
The basin maps in Sec.~IV.B show that plasticity can expand successful retrieval in the \((\alpha,M^0)\) plane. We next reduce this two-dimensional information to a single performance measure: the largest retrievable memory load at a fixed initial overlap and success threshold. This measure lets us compare different plasticity strengths directly and identify the value that maximizes basin expansion.

The feedback analysis in Sec.~IV.C suggests that the improvement of retrieval
is controlled by the strength of the total delayed feedback
\[
  \Gamma_*=\Lambda_*+K_{A,*}.
\]
In the representative successful retrieval state analyzed there, the Onsager
contribution \(\Lambda_*\) was small, whereas the plastic contribution
\(K_{A,*}\) dominated the total feedback and accurately reproduced the
converged input-output relation. Since the strength of \(\mathcal{K}_A(t,s)\) is
controlled by the plasticity parameter \(k_A\), it is natural to ask whether
there exists an optimal value of \(k_A\) for retrieval.

In this subsection, we therefore quantify the optimal plasticity strength from
the viewpoint of the basin of attraction. Unlike the fixed-point diagnostic in
Sec.~IV.C, where we ran the DMFT dynamics up to \(T=500\) in order to examine
the converged feedback, here we use the same evaluation time as in the basin
analysis of Sec.~IV.B. Namely, the final readout overlap is measured at
\[
  T_{\rm eval}=100 .
\]
For a fixed initial overlap \(M^0\) and a success threshold \(M_{\rm th}\), we
define
\[
  \alpha_{\max}(k_A;M^0,M_{\rm th})
  =
  \max\left\{
    \alpha \ \middle|\
    M^{T_{\rm eval}}(\alpha,k_A,M^0) \geq M_{\rm th}
  \right\},
\]
where the maximum is taken over the sampled values of \(\alpha\). This quantity
measures the largest memory load for which retrieval succeeds from the initial
overlap \(M^0\) at a given plasticity strength \(k_A\).

\begin{figure}[htbp]
  \centering
  \includegraphics[scale=0.8]{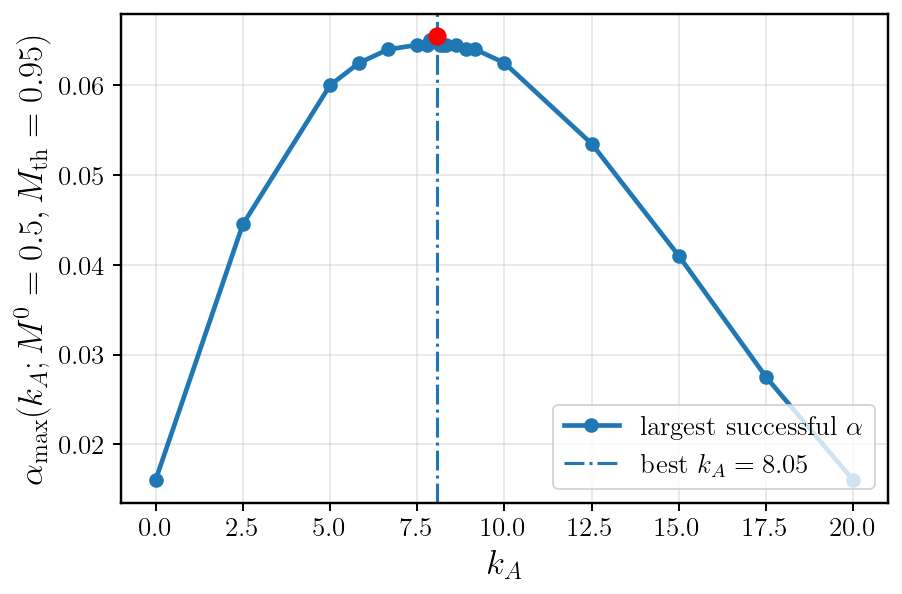}
  \caption{
    Estimation of the optimal plasticity strength for
    \(M^0=0.5\) and \(M_{\rm th}=0.95\).
    The vertical axis shows
    \(\alpha_{\max}(k_A;M^0,M_{\rm th})\), the largest memory load for which
    the final readout overlap at \(T_{\rm eval}=100\) exceeds \(M_{\rm th}\).
    Moderate plasticity enlarges the basin of attraction, whereas excessive
    plasticity reduces the retrievable region.
  }
  \label{fig:optimal_kA_curve}
\end{figure}

Figure~\ref{fig:optimal_kA_curve} shows
\(\alpha_{\max}(k_A;M^0,M_{\rm th})\) as a function of \(k_A\) for
\[
  M^0=0.5,\qquad M_{\rm th}=0.95 .
\]
The curve has a clear maximum around \(k_A\simeq 8.0\). Thus, increasing
plasticity first enlarges the retrievable region, but the improvement is not
monotonic: once \(k_A\) becomes too large, the largest retrievable memory load
decreases again.

This nonmonotonic dependence can be understood as a trade-off between two
effects of plasticity. In the moderate-\(k_A\) regime, the trajectory is first
guided toward the target memory by the static Hebbian coupling \(J_{ij}\). This is precisely the mechanism observed
in Sec.~IV.C, where the integrated plastic feedback \(K_A(t)\) dominated the
total feedback \(\Gamma(t)\) in the converged retrieval state.

On the other hand, the same mechanism becomes harmful when \(k_A\) is too large.
Since \(A^0_{ij}=0\), the first plastic update produces
\[
  A^1_{ij}
  =
  \frac{\beta}{N}\phi(x^0_i)\phi(x^0_j).
\]
Thus, if the initial cue is imperfect, the plastic coupling begins to store the
initial state itself before the neural state has sufficiently approached the
target memory. For very large \(k_A\), the field generated by this newly formed
coupling can dominate the static Hebbian retrieval field. The dynamics is then
attracted to a state reflecting the initial cue rather than to the stored target
pattern. This is a cue-imprinting failure mode.

\begin{figure}[htbp]
  \centering
  \includegraphics[width=\linewidth]{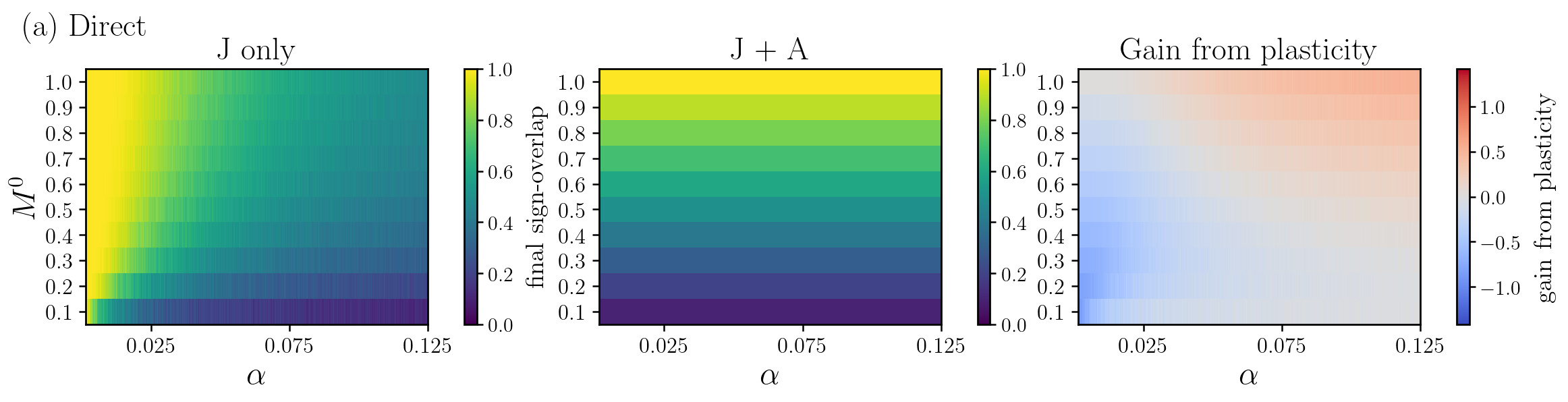}
  \caption{
    Basin of attraction under very strong plasticity, \(k_A=150\).
    This figure is a diagnostic example of the excessive-plasticity regime.
    The final overlap becomes close to the initial overlap and is almost
    independent of the memory load \(\alpha\), indicating that the plastic
    coupling imprints the initial cue rather than supporting retrieval of the
    target memory.
  }
  \label{fig:strong_plasticity}
\end{figure}

To illustrate this failure mode directly, Fig.~\ref{fig:strong_plasticity}
shows the basin heat map obtained under very strong plasticity $k_A=150$.
This value is not used in the optimization above; it is introduced only as a
diagnostic example of the excessive-plasticity regime. In this case, the final
overlap in the plastic network is nearly determined by the initial overlap and
shows little dependence on \(\alpha\). In other words, the network no longer
uses the stored Hebbian patterns efficiently. Instead, the plastic coupling
rapidly imprints the initial configuration and stabilizes it as a spurious
attractor. This observation explains the decreasing side of
Fig.~\ref{fig:optimal_kA_curve}: increasing \(k_A\) strengthens the beneficial
self-feedback, but beyond a certain point it also strengthens the undesirable
imprinting of the initial cue.

We next examine how the optimal plasticity strength depends on the initial
overlap. For each \(M^0\) and \(M_{\rm th}\), we define
\[
  k_A(M^0;M_{\rm th})
  =
  \operatorname*{arg\,max}_{k_A}
  \alpha_{\max}(k_A;M^0,M_{\rm th}),
\]
and
\[
  \alpha_{\max}(M^0;M_{\rm th})
  =
  \max_{k_A}
  \alpha_{\max}(k_A;M^0,M_{\rm th}).
\]
Here the search range is restricted to \(0\leq k_A\leq 20\) when $M_\mathrm{th}=0.95$, and \(0\leq k_A\leq 20\) when $M_\mathrm{th}=0.99$, because the purpose
is to characterize the useful retrieval-supporting regime rather than the
pathological cue-imprinting regime illustrated in
Fig.~\ref{fig:strong_plasticity}. The quantities were evaluated using the DMFT
effective process with \(10^6\) trajectories. We considered two success
thresholds,
\[
  M_{\rm th}=0.95
  \quad\text{and}\quad
  M_{\rm th}=0.999 .
\]

\begin{figure}[t]
  \centering

  \begin{subfigure}{0.48\linewidth}
    \centering
    \includegraphics[width=\linewidth]{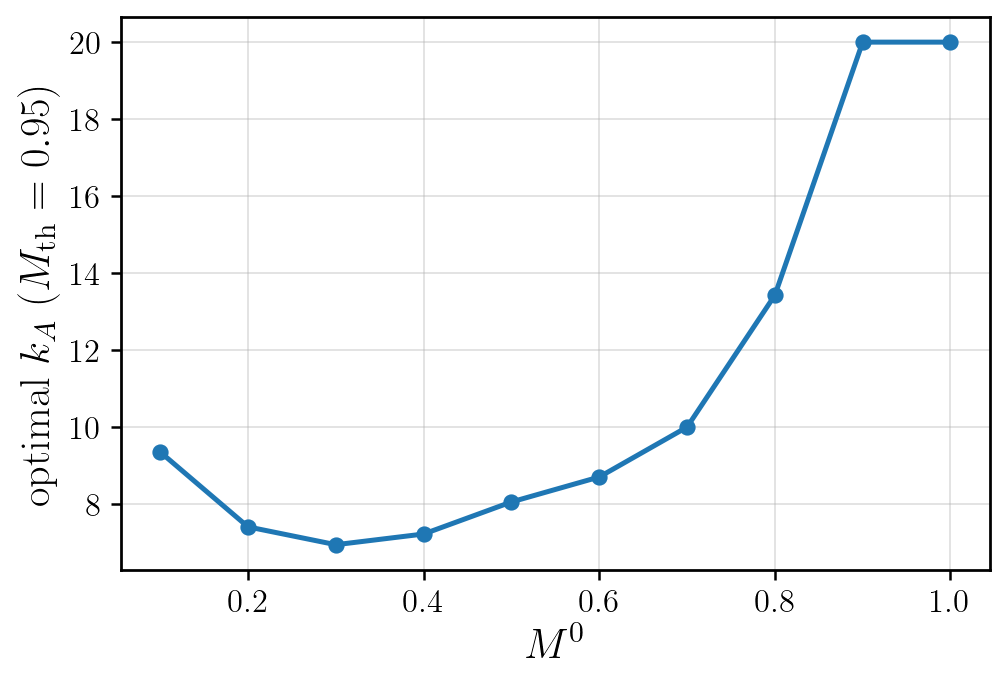}
    \caption*{(a)}
  \end{subfigure}
  \hfill
  \begin{subfigure}{0.48\linewidth}
    \centering
    \includegraphics[width=\linewidth]{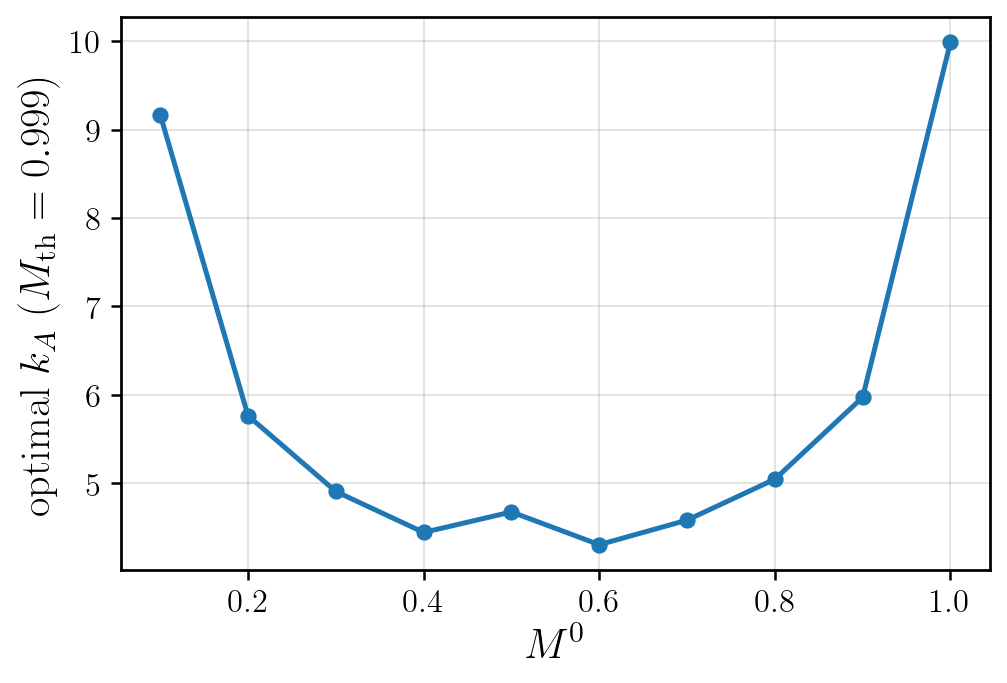}
    \caption*{(b)}
  \end{subfigure}

  \caption{
    Optimal plasticity strength \(k_A^{\rm opt}(M^0;M_{\rm th})\) as a
    function of the initial overlap \(M^0\).
    The optimization is performed over \(0\leq k_A\leq 20\) for
    (a) \(M_{\rm th}=0.95\), and \(0\leq k_A\leq 10\) for
    (b) \(M_{\rm th}=0.999\).
  }
  \label{fig:optimal_kA_vs_m0}
\end{figure}

Figure~\ref{fig:optimal_kA_vs_m0} shows
\(k_A^{\rm opt}(M^0;M_{\rm th})\) as a function of the initial overlap \(M^0\).
For both thresholds, an optimal plasticity strength exists for each \(M^0\).
The optimal value is not constant, because the role of plasticity depends on
how informative the initial cue is. When \(M^0\) is small, plasticity must be
strong enough to stabilize the trajectory after it approaches the target memory,
but excessive plasticity risks imprinting the poor initial cue. When \(M^0\) is
large, the initial cue is already close to the target pattern, so stronger
plasticity is less harmful and can more directly reinforce the correct
retrieval state. In the high-\(M^0\) region, the plotted optimum reaches the
upper search bound \(k_A=10\); this should be interpreted as the optimum within
the restricted search range, not necessarily as a global optimum over all
possible \(k_A\).

\begin{figure}[t]
  \centering

  \begin{subfigure}{0.48\linewidth}
    \centering
    \includegraphics[width=\linewidth]{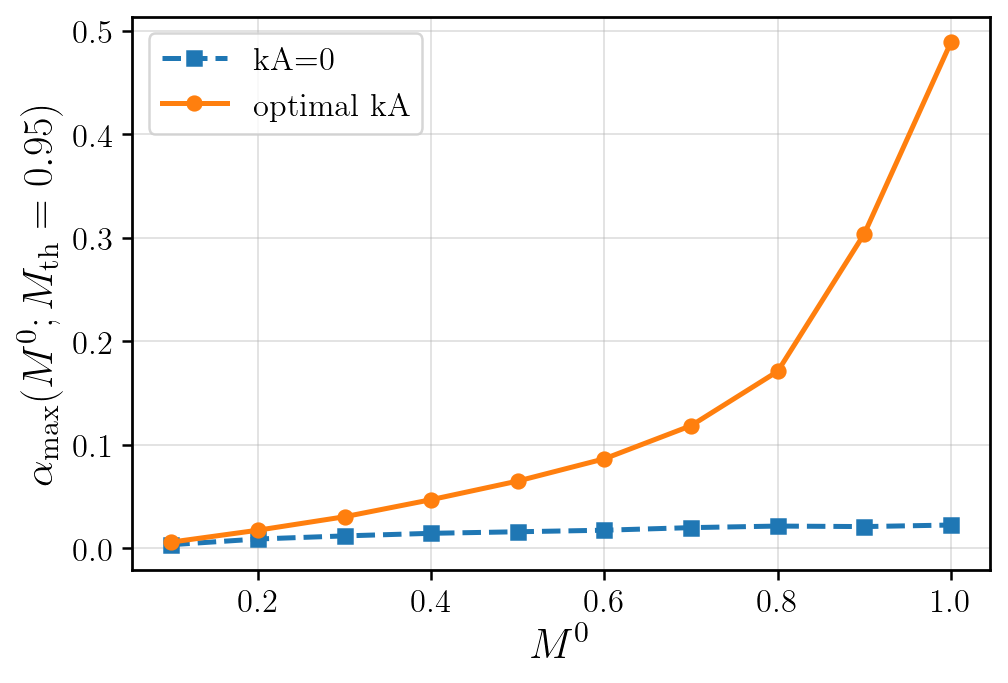}
    \caption*{(a)}
  \end{subfigure}
  \hfill
  \begin{subfigure}{0.48\linewidth}
    \centering
    \includegraphics[width=\linewidth]{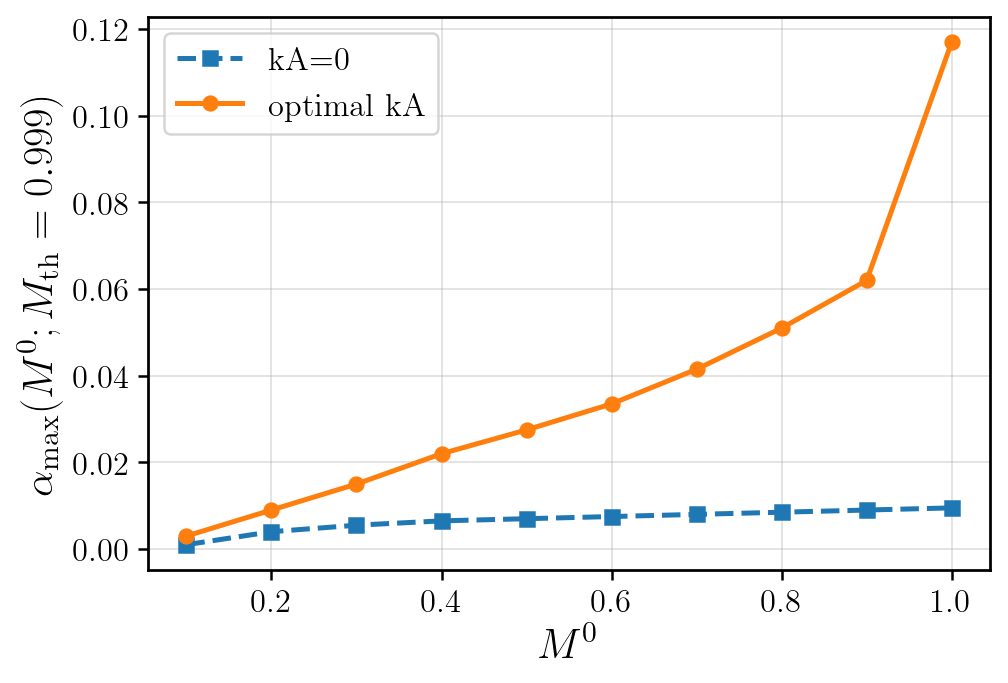}
    \caption*{(b)}
  \end{subfigure}

  \caption{
    Largest retrievable memory load as a function of the initial overlap
    \(M^0\). The optimized value
    \(\alpha_{\max}(M^0;M_{\rm th})\) is compared with the
    no-plasticity baseline.
    (a) \(M_{\rm th}=0.95\).
    (b) \(M_{\rm th}=0.999\).
    Optimizing \(k_A\) yields a substantial increase in the retrievable memory
    load for both success thresholds.
  }
  \label{fig:optimal_alpha_vs_m0}
\end{figure}

Finally, Fig.~\ref{fig:optimal_alpha_vs_m0} shows the corresponding maximum
retrievable memory load
\(\alpha_{\max}^{\rm opt}(M^0;M_{\rm th})\), together with the no-plasticity
baseline. For both success thresholds, optimizing \(k_A\) substantially
increases the retrievable memory load over a wide range of initial overlaps.
This confirms that online plasticity expands the basin of attraction not only
for a single representative initial condition, but systematically across
different values of \(M^0\).

Taken together, these results show that the effect of plasticity is
intrinsically nonmonotonic. Moderate plasticity improves retrieval by producing
the positive delayed feedback identified in Sec.~IV.C, whereas excessively
strong plasticity imprints the initial cue and creates a spurious attractor.
The optimal plasticity strength therefore reflects a trade-off between
stabilizing the target retrieval state and avoiding premature stabilization of
the imperfect initial condition.
\section{Summary and Discussion}
In this study, we analyzed a fully connected Hopfield-type associative-memory network endowed with activity-dependent plastic couplings that evolve on a timescale comparable to that of neural dynamics, thereby extending the classical fixed-synapse model of associative memory to a dynamical system in which neural activity and synaptic states are coupled~\cite{Hopfield1982,PhysRevLett.55.1530,Clark2024}. By incorporating an online Hebbian update rule into the generating-functional framework~\cite{Martin1973,Dominicis1978,Hertz_2017}, we derived a dynamical mean-field theory in the thermodynamic limit and showed that the original many-body system reduces to an effective single-site process driven by colored Gaussian noise and two delayed feedback kernels~\cite{PhysRevLett.61.259,Eissfeller_1992,Kabashima_2026}. In this formulation, the effect of plasticity appears as an additional history-dependent self-interaction determined self-consistently by the two-time correlation function. This result provides a clear macroscopic picture of how synaptic plasticity deforms retrieval dynamics in real time. More broadly, our study also connects to the viewpoint that synaptic variables should be treated not as frozen parameters but as dynamical degrees of freedom coupled to neural activity~\cite{Tsodyks1998,Clark2024}.

The comparison with Del Gaudio, Ghimenti, and Ganguli~\cite{DelGaudio2026} also helps clarify the scope of these results. Both studies consider Hopfield associative memories with an online Hebbian plastic matrix that coevolves with neural activity, and both show that a static fixed-point analysis alone misses an important dynamical retrieval mechanism. In their terminology, short-term plasticity stabilizes transient retrieval through a trampoline mechanism. In the present discrete-time formulation, the same physical effect appears as a plasticity-induced retarded self-feedback that stabilizes the retrieval branch selected by the trajectory. The main difference lies in the quantitative emphasis. Ref.~\cite{DelGaudio2026} focused on the continuous-time formulation, the relation between static cavity theory and DMFT, and the optimal plasticity timescale. Here, we used generating-functional DMFT to quantify basin expansion as a function of initial overlap and memory load, and to characterize the optimal plasticity strength. This analysis also exposes the opposite side of the same feedback mechanism: when plasticity is too strong, the initial cue can be imprinted before the trajectory reaches the stored target pattern, producing a spurious cue-dependent attractor.

The numerical results showed that, in the parameter region examined in this paper, the predictions of DMFT agree very well with direct simulations. Moreover, moderate plasticity improves retrieval performance in two complementary senses: it enlarges the basin of attraction and increases the maximum memory load for which reliable retrieval is possible. These findings suggest that online synaptic modification can function not only as a mechanism for learning and consolidation but also as a dynamical process that directly supports retrieval itself. This conclusion is consistent with previous studies of associative-memory networks with dynamic synapses, where synaptic depression, synaptic facilitation, and related dynamical synaptic variables were shown to alter retrieval phases, memory capacity, attractor switching, and stability~\cite{Pantic2002,Mejias2009,Otsubo2010,Katori2013}.

At the same time, our analysis clarified that the effect of plasticity is not monotonic. When the plasticity strength becomes too large, the synaptic matrix rapidly imprints the initial state, and the network is driven toward a spurious attractor reflecting this transient initial configuration rather than the target memory. Such spurious attractors are one of the defining features of Hopfield-type associative-memory models~\cite{Hopfield1982,PhysRevLett.55.1530}. As a result, retrieval performance deteriorates, and an optimal plasticity strength emerges for each initial overlap and success threshold. This trade-off between supporting retrieval and generating false memories appears to be one of the central dynamical consequences of online plasticity in associative-memory networks. It is also conceptually related to the parameter-dependent changes in stability found in associative-memory models with dynamic synapses~\cite{Mejias2009,Katori2013}.

The present results also clarify the relation between online Hebbian plasticity and other adaptive extensions of dense associative-memory models proposed in recent years. In recent gated associative-memory models, activity-dependent self-adaptive gating has been analyzed by many-body simulations and DMFT, and has been shown to reorganize attractor structures, extend high-overlap retrieval, and generate multistability~\cite{article}. Although such gating variables are not Hebbian synaptic traces themselves, they are conceptually close to our model in that additional adaptive degrees of freedom deform the retrieval landscape on the fly. In another complementary direction, recent studies of ongoing associative synaptic plasticity have shown that coupling between neural activity and synaptic dynamics can generate dynamically persistent activity~\cite{Wakhloo2025}. Compared with these studies, the present work focuses on the case in which an online Hebbian trace is added to the classical Hopfield memory-embedding rule, and shows that the same real-time plastic mechanism can both improve retrieval and create failure modes by imprinting transient states.

For future work, it will be important on the theoretical side to characterize analytically the optimal plasticity strength and the stability boundaries of the retrieval state. It will also be necessary to examine whether similar conclusions hold for correlated patterns, sparse networks, or other plasticity rules such as short-term depression, short-term facilitation, and spike-timing-dependent plasticity~\cite{Tsodyks1997,Zucker2002,Tsodyks1998}. On the applied side, extending the framework to sequence retrieval, temporally structured inputs, and more biologically realistic circuit architectures may lead to a deeper understanding of the conditions under which dynamic synapses facilitate memory computation and those under which they interfere with it. As a further natural extension, one may combine the Hebbian plasticity kernel studied here with additional adaptive variables such as gating variables or neuromodulatory signals~\cite{article}, or investigate whether similar mechanisms generate temporally structured persistent memories in coupled neural-activity--synaptic systems~\cite{Wakhloo2025}.

\section*{acknowledgments}
YH gratefully acknowledges financial support from The RIKEN Junior Research Associate (JRA) program.
This work was supported by the Forefront Physics and Mathematics Program to Drive Transformation
(FoPM), the World-leading Innovative Graduate Study (WINGS) Program, the University of Tokyo
(YH), and MEXT/JSPS KAKENHI Grant Number 22H05117
and JSPS KAKENHI Grant Number 26K02981 (YK).

\clearpage
\appendix

\section{Average over the Quenched Patterns}
In this appendix, we give the details of the quenched average over the nonretrieved patterns. We first fix the retrieved pattern to $\xi_i^1=1$ by a gauge transformation, decompose the Hopfield field into a signal term and a crosstalk-noise term, and then evaluate the contribution of the remaining patterns by a Gaussian approximation in the thermodynamic limit.

Next, note that the activation function is odd, namely $\phi(x)=-\phi(-x)$. Then, by symmetry and a gauge transformation, the retrieved pattern with $\mu=1$ can be treated~\cite{PhysRevLett.55.1530, Kabashima_2026} as
\begin{align}
  \xi^1_i=1\quad (i=1,\ldots,N)
\end{align}
With this choice, the local field generated by the Hopfield coupling can be decomposed into a signal contribution from the retrieved pattern and a crosstalk-noise contribution from the remaining patterns as follows~\cite{PhysRevLett.55.1530, Kabashima_2026}.
\begin{align}
  \sum_{j\neq i}J_{ij}\phi(t_j)&=\frac{g}{N}\sum_{j}\phi(t_j)+\frac{g}{N}\sum_{\mu=2}^P \xi^\mu_i\sum_{j=1}^N \xi^\mu_j \phi(t_j)-\frac{gP}{N}\phi(t_i)\\
  \label{eq:field_decomposition}
  &=gm^t+\frac{g}{\sqrt{N}}\sum_{\mu=2}^P \xi^\mu_iu^t_\mu-\alpha g\phi(x^t_i).
\end{align}
where
\begin{align}
  m^t=\frac{1}{N}\sum_{i=1}^N \phi(x^t_i),\quad u^t_\mu=\frac{1}{\sqrt{N}}\sum_{i=1}^N\xi^\mu_i \phi(x^t_i)
\end{align}
were defined. The Fourier representation of the delta functions in Eq.~(\ref{eq:generating_functional}) is
\begin{align}
  \label{eq:delta_fourier}
  \delta\!\left(x_i^{t+1}-\mathcal{F}_i^t[x_i^{0:s\le t}]\right)
  =\int\frac{\dd{\hat{x}_i^t}}{2\pi}
  \exp\!\left[-i\hat{x}_i^t\left(x_i^{t+1}-\mathcal{F}_i^t[x_i^{0:s\le t}]\right)\right].
\end{align}
Using Eqs.~(\ref{eq:generating_functional}), (\ref{eq:delta_fourier}), and (\ref{eq:field_decomposition}), the average over the random patterns reduces to the evaluation of
\begin{align}
  \mathbb{E}_\xi\qty[\exp\qty(\gamma g\sum_{\mu=2}^P \sum_{t=0}^{T-1}v^t_\mu u^t_\mu)]
\end{align}
where we set
\begin{align}
  v^t_\mu=\frac{1}{\sqrt{N}}\sum_{i=1}^N\xi_i^\mu (i\hat{x}^t_i)
\end{align}
\par
For fixed trajectories $x$ and $\hat{x}$, the variables $u$ and $v$ can, in the large system limit, be treated as jointly Gaussian random variables with zero mean by the central limit theorem~\cite{Dominicis1978, Hertz_2017, Zou_2024}. Their covariances are given by
\begin{align}
  &\mathbb{E}_\xi\qty[u^t_\mu u^s_\nu]=\delta_{\mu\nu}Q(t,s),\quad Q(t,s)=\frac{1}{N}\sum_{i=1}^N\sum_{i=1}^N\phi(x^t_i)\phi(x^s_i),\\
  &\mathbb{E}_\xi\qty[v^t_\mu v^s_\nu]=\delta_{\mu\nu}R(t,s),\quad R(t,s)=\frac{1}{N}\sum_{i=1}^N (i\hat{x}^t_i)(i\hat{x}^s_i),\\
  &\mathbb{E}_\xi\qty[u^t_\mu v^s_\nu]=\delta_{\mu\nu}S(t,s),\quad S(t,s)=\frac{1}{N}\sum_{i=1}^N\phi(x^t_i)(i\hat{x}^s_i)
\end{align}
Using this, for a single pattern with $\mu\geq 2$ we obtain
\begin{align}
  \mathcal{I}_\mu&=\mathbb{E}_\xi\qty[\exp\qty(\gamma g\sum_{t=0}^{T-1}v^t_\mu u^t_\mu)]\nonumber\\
  \label{eq:gaussian_average}
  &=\qty\bigg[\det Q\det\qty\Big{(I-\gamma gS)^\mathsf{T}Q^{-1}(I-\gamma gS)-\gamma^2 g^2 R}]^{-1/2}
\end{align}
Hence, after averaging over the $P-1\simeq \alpha N$ patterns, the factor in Eq.~(\ref{eq:gaussian_average}) contributes to the generating functional raised to the power $\alpha N$.\par

\section{Saddle-Point Evaluation and Effective Single-Site Process}
In this appendix, we introduce the order parameters and their conjugate variables, perform the Hubbard--Stratonovich transformation, and evaluate the averaged generating functional by the saddle-point method. This yields the effective single-site process, the self-consistency equations, and the update rule for the response function used in the main text.

Here we keep the conjugate variables that do not appear explicitly in the main text and spell out in more detail how the averaged generating functional reduces to a single-site process. The key point is that $R=0$ is a consequence of the saddle-point equations rather than an assumption at the outset, and that $\hat{R}$ must be retained at intermediate stages because it generates the covariance of the effective Gaussian noise through the Hubbard--Stratonovich decoupling.

Collecting the order parameters and conjugate variables as
\begin{align}
  \Omega=\{m,Q,R,S\},
  \qquad
  \hat{\Omega}=\{\hat{m},\hat{Q},\hat{R},\hat{S}\},
\end{align}
the disorder-averaged generating functional obtained after Eq.~(\ref{eq:gaussian_average}) can be written as
\begin{align}
  \mathbb{E}_{\xi}[Z(\theta)]
  =\int d\Omega\,d\hat{\Omega}\,
  \exp\qty(N\Psi[\Omega,\hat{\Omega}])
  \prod_{i=1}^{N}Z_i[\theta_i;\Omega,\hat{\Omega}],
\end{align}
with
\begin{align}
  Z_i[\theta_i;\Omega,\hat{\Omega}]
  &=\int \mathcal{D}x_i\,\mathcal{D}\hat{x}_i\,P_0(x_i^0)
  \exp\qty(\Phi_i[x_i,\hat{x}_i;\Omega,\hat{\Omega}]),
\end{align}
\begin{align}
  \Phi_i
  =&-\sum_{t=0}^{T-1}i\hat{x}_i^t\Biggl[
    x_i^{t+1}-x_i^t
    -\gamma\Bigl(
      -x_i^t+gm^t-\alpha g\phi(x_i^t)
      +\sum_{s<t}\mathcal{K}_A(t,s)\phi(x_i^s)
    \Bigr)
  \Biggr]
  +i\sum_{t=1}^{T}\theta_i^tx_i^t\nonumber\\
  &-\sum_{t=0}^{T-1}\hat{m}^t\phi(x_i^t)
  -\frac{1}{2}\sum_{t,s=0}^{T-1}\hat{Q}(t,s)\phi(x_i^t)\phi(x_i^s)\nonumber\\
  &-\sum_{t,s=0}^{T-1}\hat{S}(t,s)\phi(x_i^t)(i\hat{x}_i^s)
  -\frac{1}{2}\sum_{t,s=0}^{T-1}(i\hat{x}_i^t)\hat{R}(t,s)(i\hat{x}_i^s),
\end{align}
and
\begin{align}
  \Psi[\Omega,\hat{\Omega}]
  =&-\frac{\alpha}{2}\ln\det Q
  -\frac{\alpha}{2}\ln\det M
  +\sum_{t=0}^{T-1}\hat{m}^tm^t\nonumber\\
  &+\frac{1}{2}\Tr(\hat{Q}Q)
  +\Tr(\hat{S}S)
  +\frac{1}{2}\Tr(\hat{R}R),
\end{align}
where
\begin{align}
  M=(I-\gamma gS)^\mathsf{T}Q^{-1}(I-\gamma gS)-\gamma^2g^2R.
\end{align}

$\hat{R}$ multiplies the quadratic form of the response field $i\hat{x}$, and it is this term that is decoupled by the Hubbard--Stratonovich transformation. By contrast, the $\hat{S}$ term is only linear in $i\hat{x}$, so it does not require a separate Gaussian decoupling and is instead absorbed into the retarded self-interaction kernel.

Specifically, for each site we use the identity
\begin{align}
  &\exp\qty[-\frac{1}{2}\sum_{t,s=0}^{T-1}(i\hat{x}_i^t)\hat{R}(t,s)(i\hat{x}_i^s)]\nonumber\\
  &=\int \mathcal{D}z_i\,
  \exp\qty[-\frac{1}{2}\sum_{t=0}^{T-1}(z_i^t)^2
  +\sum_{t=0}^{T-1}i\hat{x}_i^t\qty(\sqrt{-\hat{R}}\,z_i)^t],
\end{align}
and define $\eta_i=\sqrt{-\gamma^{-2}\hat{R}}\,z_i$. Then the $\hat{R}$ term becomes an additive Gaussian field in the local equation, and the local action is linear in $i\hat{x}_i$. At this stage the noise covariance is still written in terms of the conjugate variable $\hat{R}$.

We next introduce the response function
\begin{align}
  G(t,s)=\gamma S(t,s)
\end{align}
Taking the saddle point, one first obtains from stationarity with respect to $R$
\begin{align}
  \frac{\partial \Psi}{\partial R}=0
  \qquad\Longrightarrow\qquad
  \hat{R}=-\alpha\gamma^2g^2M^{-1}.
\end{align}
The normalized generating functional selects the saddle with $R=0$, but this is imposed only after the above derivative has been taken. Substituting $R=0$ into the expression for $\hat{R}$ gives the covariance of the effective Gaussian noise,
\begin{align}
  C=-\gamma^{-2}\hat{R}
  =\alpha g^2(I-gG)^{-1}Q(I-gG)^{-\mathsf{T}}.
\end{align}

Similarly, stationarity with respect to $Q$ yields $\hat{Q}=0$ once $R=0$ is used, while stationarity with respect to $S$ gives
\begin{align}
  \hat{S}=-\alpha\gamma gQ^{-1}(I-\gamma gS)M^{-1}
  =-\alpha\gamma g(I-gG)^{-\mathsf{T}}.
\end{align}
Combining this contribution with the explicit self-interaction term $-\alpha g\phi(x_i^t)$ produces the Onsager-reaction kernel
\begin{align}
  \Lambda
  =-\alpha gI-\gamma^{-1}\hat{S}^{\mathsf{T}}
  =\alpha g^2G(I-gG)^{-1}.
\end{align}

After inserting these saddle-point relations back into the site-local action, the disorder-averaged generating functional becomes the generating functional of $N$ identical noninteracting stochastic processes~\cite{PhysRevLett.61.259, Eissfeller_1992, Hertz_2017, Zou_2024}. Concretely, the original many-body interacting system reduces to the following effective single-site process~\cite{PhysRevLett.61.259,Clark2024,Kabashima_2026}.
\begin{align}
  x^{t+1}=x^t+\gamma\qty(-x^t+gm^t+\eta^t+\sum_{s<t}\Lambda(t,s)\phi(x^s)+\sum_{s<t}\mathcal{K}_A(t,s)\phi(x^s))
\end{align}
Here, $\eta^t$ is a colored Gaussian noise with zero mean satisfying
\begin{align}
  \expval{\eta^t\eta^s}=C(t,s)
\end{align}
where $\expval{\cdots}$ denotes the average over this effective single-site process.\par
The self-interaction term $-\alpha g \phi(x^t_i)$ that appeared in Eq.~(\ref{eq:field_decomposition}) does not appear explicitly in Eq.~(\ref{eq:effective_single_site}). Instead, after being combined with the term arising from the pattern average, it is absorbed into the causal delayed-feedback kernel $\Lambda(t,s)$~\cite{PhysRevLett.55.1530, Hertz_2017, Kabashima_2026}. Therefore, $\Lambda(t,s)$ has nonzero components only for $s\leq t$.\par
The self-consistency equations are given by
\begin{align}
  m^t=\expval{\phi(x^t)},\quad Q(t,s)=\expval{\phi(x^t)\phi(x^s)}\\
  \mathcal{K}_A(t,s)=\beta\rho^{t-1-s} Q(t,s),\quad s<t\\
  C(t,s)=\alpha g^2(I-gG)^{-1}Q(I-gG)^{-\mathsf{T}}(t,s)\\
  \Lambda(t,s)=\alpha g^2 G(I-gG)^{-1}(t,s)
\end{align}
Here, $G$ is a lower-triangular matrix satisfying the causal condition $G(t,s)=0$ for $s>t$, and in numerical calculations it is updated~\cite{Eissfeller_1992, Zou_2024, Kabashima_2026} as
\begin{align}
  G(t,s)=
  \begin{cases}
    \displaystyle\sum_{w=0}^{T-1}\expval{\phi(x^t)\eta^w}\qty[C^{-1}_{<t}(w,s)], &s<t,\\
    0,&s\geq t.
  \end{cases}
\end{align}

Equations (\ref{eq:effective_single_site})--(\ref{eq:response_function_update}) constitute the DMFT equations for the fully connected associative-memory network with plasticity. Crosstalk noise from the nonretrieved patterns appears both as the fluctuation term $\eta^t$ and as the Onsager reaction term $\Lambda(t,s)$, which acts as self-feedback. In addition, the effect of plasticity enters the effective single-site process as another delayed self-feedback term $\mathcal{K}_A(t,s)$. Therefore, the macroscopic dynamics of the fully connected associative-memory model with plasticity can be described in closed form as an effective single-site process with colored Gaussian noise and two kinds of delayed feedback~\cite{PhysRevLett.61.259, Clark2024, Kabashima_2026}.\par

\end{document}